\documentclass[aos,MSNbibl,seceqn,rotating,dvips]{arximspdf}
\usepackage{dcolumn}
\usepackage{graphicx}

\doi{10.1214/14-AOS1236} %kopijuoti is PTS
\volume{42}
\issue{5}
\pubyear{2014}
\firstpage{1819}
\lastpage{1849}

\makeatletter
\newcolumntype{d}[1]{D{.}{.}{#1}}
\newtheorem{teo}{Theorem}[section]
\newproclaim{rmk}{Remark}
\newproclaim{ass}{Assumption}
\newproclaim{SP}{Sparsity property}
\newproclaim{OP}{Oracle property}
\newcommand{\Cvar}{\mathit{Cvar}}
\newcommand{\PLS}{\mathrm{PLS}}
\newcommand{\cov}{\operatorname{Cov}}
\newcommand{\RP}{\mathrm{P}}
\newcommand{\RE}{\mathrm{E}}
\newcommand{\var}{\operatorname{Var}}
\newcommand{\town}{\mathbf{W}}
\newcommand{\0}{\mathbf{0}}
\newcommand{\1}{\mathbf{1}}
\newcommand{\ba}{\mathbf{a}}
\newcommand{\bI}{\mathbf{I}}
\newcommand{\bB}{\mathbf{B}}
\newcommand{\bW}{\mathbf{W}}
\newcommand{\bG}{\mathbf{G}}
\newcommand{\bS}{\mathbf{S}}
\newcommand{\bV}{\mathbf{V}}
\newcommand{\bX}{\mathbf{X}}
\newcommand{\bY}{\mathbf{Y}}
\newcommand{\bT}{\mathbf{T}}
\bmdefine{\bgamma}{\gamma}
\bmdefine{\bdelta}{\delta}
\bmdefine{\bDelta}{\Delta}
\bmdefine{\bLambda}{\Lambda}
\bmdefine{\bOmega}{\Omega}
\bmdefine{\bSigma}{\Sigma}
\bmdefine{\balpha}{\alpha}
\bmdefine{\bbeta}{\beta}
\bmdefine{\bphi}{\phi}
\bmdefine{\bxi}{\xi}
\bmdefine{\bta}{\eta}
\makeatother

\begin{document}
\begin{frontmatter}

\title{Nonparametric independence screening and structure
identification for ultra-high dimensional~longitudinal data}
\runtitle{Ultra-high dimensional longitudinal data}

\begin{aug}
\author[a]{\fnms{Ming-Yen} \snm{Cheng}\corref{}\thanksref{t1}\ead[label=e1]{cheng@math.ntu.edu.tw}},
\author[b]{\fnms{Toshio} \snm{Honda}\thanksref{t2}\ead[label=e2]{t.honda@r.hit-u.ac.jp}},
\author[c]{\fnms{Jialiang} \snm{Li}\thanksref{t3}\ead[label=e3]{stalj@nus.edu.sg}}
\and\break
\author[d]{\fnms{Heng} \snm{Peng}\thanksref{t4}\ead[label=e4]{hpeng@math.hkbu.edu.hk}}
\runauthor{Cheng, Honda, Li and Peng}
\affiliation{National Taiwan University,
Hitotsubashi University,\\
National University of Singapore
and Hong Kong Baptist University}
\address[a]{M.-Y. Cheng\\
Department of Mathematics\\
National Taiwan University\\
Taipei 106\\
Taiwan\\
\printead{e1}}
\address[b]{T. Honda\\
Graduate School of Economics\\
Hitotsubashi University\\
Kunitachi, Tokyo 186-8601\\
Japan\\
\printead{e2}}
\address[c]{J. Li\\
Department of Statistics \& Applied Probability\\
National University of Singapore\\
Singapore 117546\\
\printead{e3}}
\address[d]{H. Peng\\
Department of Mathematics\\
Hong Kong Baptist University\hspace*{3pt}\\
Kowloon\\
Hong Kong\\
China\\
\printead{e4}}
\end{aug}
\thankstext{t1}{Supported in part by the National Science Council
Grants NSC97-2118-M-002-001-MY3 and NSC101-2118-M-002-001-MY3,
and the Mathematics Division, National Center of Theoretical Sciences
(Taipei Office).}
\thankstext{t2}{Supported by the JSPS Grant-in-Aids for Scientific
Research (A) 24243031 and (C) 25400197.}
\thankstext{t3}{Supported by Grants AcRF R-155-000-130-112 and
NMRC/CBRG/0014/2012.}
\thankstext{t4}{Supported by CERG Grants from the Hong Kong Research
Grants Council
(HKBU 201610 and HKBU 202012), FRG Grants from Hong Kong Baptist
University (FRG2/11-12/130 and FRG2/12-13/077), and a Grant from NSFC
(11271094).}

% HISTORY:
\received{\smonth{8} \syear{2013}}
\revised{\smonth{5} \syear{2014}}

% ABSTRACT
%
\begin{abstract}
Ultra-high dimensional longitudinal data are increasingly common
and the analysis is challenging both theoretically and methodologically.
We offer a new automatic procedure for finding
a sparse semivarying coefficient model, which is widely accepted for
longitudinal data analysis.
Our proposed method first reduces the number of covariates to a moderate
order by employing a screening procedure, and then identifies both
the varying and \mbox{constant} coefficients using a group SCAD estimator,
which is subsequently refined by
accounting for the within-subject correlation.
The screening procedure is based on working independence and B-spline
marginal models.
Under weaker conditions than those in the literature, we show that with
high probability only
irrelevant variables will be screened out, and the number of selected
variables can be bounded by a moderate order. This allows the desirable
sparsity and oracle properties of the subsequent structure
identification step. %It also marks the significance of our theory and
%methodology,
Note that existing methods require some kind of iterative screening in
order to achieve this,
thus they demand heavy computational effort and consistency is not
guaranteed. %We prove that our group SCAD estimator detects the
%constant and varying effects simultaneously.
The refined semivarying coefficient model employs
profile least squares, local linear smoothing and nonparametric
covariance estimation,
and is semiparametric efficient.
We also suggest ways to implement
the proposed methods, and to select the tuning parameters. An extensive
simulation study
is summarized to demonstrate its finite sample performance and the
yeast cell cycle data is analyzed.
\end{abstract}

% KEYWORDS
\end{frontmatter}

%s1 #&#
\section{Introduction}
\label{secsetup}

Longitudinal data arise in many modern scientific fields, including
finance, genetics, medicine and so on.
Specifically, we consider observing independent\vspace*{1pt} realizations of a
scalar response process $y(t)$ and a $p$-dimensional covariate process
$\mathbf{x}(t)=(x^{( 1 )}( t ), \ldots, x^{( p )}( t ))^T$ at $t=t_1,
\ldots, t_m$, where $t_1, \ldots, t_m$, independent of $ \mathbf
{x}(t)$, are
i.i.d. with density $f_T(t)$ satisfying $C_1\le f_T(t) \le C_2$.
In this paper,
$C, C_1, C_2, \ldots$ are positive generic constants. %whose values
%vary from place to place.
%We write $\mathbf{t}\equiv(t_1, \ldots, t_m)$. %Assume that ${\mathbf t}
There exist various parametric, nonparametric and semiparametric models
for regressing $y(t)$ on ${\mathbf x}(t)$ %see for example
\mbox{\cite{FW2008,ZFS2009}}. %\cite{YL2013}, and references therein.
Among the three categories, the semiparametric approach, in particular
varying coefficient models, is in general preferred to the other two.
Parametric models are efficient if correctly specified, but can be
seriously biased otherwise. While nonparametric approach avoids this
problem, the curse of dimensionality issue arises.% even for small $p$.

Consider the varying coefficient model, which can capture the dynamical
impacts of the covariates on the response variable, given as
%
%e1.1 #&#
\begin{equation}
y(t) = \beta_0 (t) + \sum_{k=1}^p
x^{( k )}( t )\beta_k(t) + \varepsilon(t),\qquad t
\in[0,1],\label{eqnh101}
\end{equation}
where $\beta_0(t), \beta_1(t), \ldots, \beta_p(t)$ are the unknown
varying coefficients and $\varepsilon(t)$ is an error process with $\RE
\{ \varepsilon(t) | {\mathbf x}(t)\} =0$.
For a generic real-valued function $g$, % on $[0,1]$,
write $g({\mathbf t})=(g(t_{1}),\ldots,g(t_{m}))$, where ${\mathbf t}\equiv
(t_1, \ldots, t_m)$.
Suppose we are given $n$ independent observations on $(y({\mathbf t}), {\mathbf
x}({\mathbf t}))$: for the $i$th subject, we observe
$y_i(t)$ and ${\mathbf x}_i(t) = (x_{ i }^{( 1 )}( t ), \ldots, x_{ i }^{(
p )}( t ))^T$
at $t=t_{i1}, \ldots, t_{im_i}$. %, for $ i=1, \ldots, n$.
Here, $m_i$ can be random, but is uniformly bounded
and independent of ${\mathbf x}_i(t)$.
Writing ${\mathbf t}_i = (t_{i1}, \ldots, t_{im_i})^T$, we have $(y_i({\mathbf
t}_i),{\mathbf x}_i({\mathbf t}_i))$ where %$y_i(\bt_i)=(y_i(t_{i1}),
${\mathbf x}_i({\mathbf t}_i)= ({\mathbf x}_i(t_{i1}), \ldots, {\mathbf
x}_i(t_{im_i}))$ is a $p\times m_i$ random matrix. %, $i=1,\ldots,n$.
Based on model (\ref{eqnh101}), we have %the following varying
%coefficient model for the data
%
%e1.2 #&#
\begin{equation}
\label{modelfull} y_i({\mathbf t}_i) = \beta_0 ({
\mathbf t}_i) + \sum_{k=1}^p
x_{ i }^{( k )}( {\mathbf t}_i )\beta_k({\mathbf
t}_i)^T + \varepsilon_i({\mathbf
t}_i),\qquad i=1,\ldots,n,
\end{equation}
where %$x^{(k)}_i(\bt_i)=(x^{(k)}_i(t_{i1}),
$\varepsilon_i ({\mathbf t}_i ) = (\varepsilon_i(t_{i1}), \ldots,\varepsilon
_i(t_{im_i}))$ is the error process in the $i$th subject. %$(y_i(
%Note that $y_i(\bt_i)$, $\xb{i}{k}{\bt_i}$, $\beta_k (\bt_i)$ and $

As technology evolves rapidly over the recent decades, high-dimensional
longitudinal data have become commonly encountered, and the analysis
poses new challenges from methodological, theoretical and computational aspects.
When $p$ is large, it is often the case that many of the covariates are
irrelevant. Under such circumstances, including the irrelevant
variables in the model would create undesirable identifiability and
estimation instability problems, and variable selection is a natural
way to address the challenges.
In parametric regression for i.i.d. data, popular tools for this
purpose include the
SCAD \cite{FL2001}, Lasso
\cite{Tibshirani1996}, adaptive Lasso \cite{Zou2006}
and group Lasso \cite{MGB2008,YL2006} estimators. %, which can select
%variables and estimate the parameters simultaneously.
These ideas have been adopted to select important variables in varying
coefficient models for i.i.d. data, that is, $m_1=\cdots m_n=1$,
\cite{XQ2012}. %. Moreover, a significant number of papers have considered
%the variable selection problem for model (\ref{modelfull}).
For longitudinal data, when $p$ is fixed, group SCAD penalized \mbox{B-}spline
methods were studied in \cite{WLH2008} and
\cite{NP2010}, and regularized\vspace*{1pt} P-spline methods were considered in
\cite{AGV2012}. When $p$ diverges %to $\infty$
and $p=o(n^{2/5})$, where $n$ is the sample size, \cite{WHL2011} and
\cite{AGL2012} examined adaptive group Lasso estimators.

However, it occurs often in today's longitudinal studies that $p$ is
very large. An example we will investigate in Section~\ref{yeast} is
the famous yeast cell cycle data set, which consists of gene expression
measurements at different time points during the cell cycle \cite
{Spellman1998}. In this dataset, $p=96$ and $n= 297$, thus $p$ is much
larger than $n^{2/5}\approx10$.
Under such circumstances, there is no guarantee that existing variable
selection procedures can find the relevant variables consistently. We
consider the more general ultra-high dimensional case where $p$ can be
lager than $n$.
Our idea is first reducing the dimensionality to a moderate order by
employing some screening procedure, and then selecting variables using
a group SCAD estimator, which possesses the desirable sparsity and
oracle properties.
In parametric settings, existing screening methods include the sure
independence screening procedures \mbox{\cite{FL2008,FS2010}}, the rank
correlation screening procedure \cite{Lipengzhangzhu2012} and
others. In semiparametric settings, screening procedures have been
considered for additive and varying coefficient models when the data
are i.i.d. \cite{FFS2011,FMD2013,LLW2013}.
In the present setup, we adopt the nonparametric independence screening
(NIS) idea in \cite{FFS2011}. Moreover, the covariance structure of
$\varepsilon(t)$ is unknown in general, and it is infeasible to estimate
it at this stage. We base our NIS procedure on a working independence
structure. Intuitively, this approach is expected to work since the
coefficient estimators based on working independence achieve the same
convergence rate as that based on the true covariance structure.

Under weaker conditions than in the literature, our NIS step can
effectively cut the dimensionality down to a moderate order. %,
%$q=o(n^{2/5}/\sqrt{\log n})$ say.
Writing as $x^{(1)},\ldots,\break x^{(q)}$ the remaining variables after the
NIS step, we now reduce the full varying coefficient model (\ref
{modelfull}) to the following lower-dimensional one:
%
%e1.3 #&#
\begin{equation}
\label{modelnis0} y_i({\mathbf t}_i) = \beta_0 ({
\mathbf t}_i) + \sum_{k=1}^q
x_{ i }^{( k )}( {\mathbf t}_i )\beta_k({\mathbf
t}_i)^T + \varepsilon_i({\mathbf
t}_i),\qquad i=1,\ldots,n.
\end{equation}
Under appropriate smoothness assumptions, we can estimate the unknown
coefficient functions %$\beta_k(t)$, $k=0,1,\ldots,q$,
in model (\ref{modelnis0}) using %an equi-spaced \mbox{B-}spline basis of
%order more than or equal to $2$ on $[0,1]$
\mbox{B-}spline smoothing \cite{Schumaker2007}.
However, the dimension $q$ may be still too large for the modeling
purpose, thus it is preferable to further select among these $q$
variables the significant ones. Noticeably, we can proceed directly
with variable selection as we show that $q$ can be controlled at
$o(n^{2/5}/\sqrt{\log n})$ after the NIS step, while existing methods
require some sort of iterative screening to achieve similar goals
\cite{FL2008,FS2010}. We choose the SCAD penalty in both of the variable
screening and selection steps because it enjoys a faster convergence
rate than the Lasso $L_1$ penalty when the dimension is very large
\cite{FL2013}.

Besides variable screening and variable selection, we pay attention to
the structure identification problem. That is, some of the important
variables may simply have constant effects. Identifying the nonzero
constant coefficients is an important issue because treating a constant
coefficient as varying will yield a slower convergence rate than $\sqrt{n}$.
When $p$ is fixed, significant effort has been devoted to address this
problem in varying coefficient models for both i.i.d. and longitudinal
data \cite{XZT2004,ZFS2009,taoxia2011}. %For longitudinal data,
%in the partially linear model i.e. all variables have constant
%effects, and \cite{ZX2012} used partial group SCAD to select variables
%with varying coefficients assuming the variable with constant effects
%are given. Clearly, these two papers did not consider the structure
%identification problem.
In addition, structure identification was considered for partially
linear additive models by \cite{ZCL2011} and for Cox proportional
hazard models with varying coefficients by \cite{YH2012,LLL2013}.
To achieve simultaneous variable selection and structure
identification, we construct a group SCAD penalty to penalize both
spurious nonconstant effects and spurious nonzero effects. After this
step, we further reduce the varying coefficient model (\ref
{modelnis0}) to the following semivarying coefficient model:
%
%e1.4 #&#
\begin{equation}
y_i({\mathbf t}_{i}) = \beta_0({\mathbf
t}_i) + \sum_{k=1}^{s_1}
x_{ i }^{( k
)}( {\mathbf t}_i )\beta_k +
\sum_{k=s_1+1}^{s} x_{ i }^{( k )}(
{\mathbf t}_i )\beta_k({\mathbf t}_i)^T
+ \varepsilon_i({\mathbf t}_{i}), \label{modelsemivarying0}
\end{equation}
$i=1,\ldots,n$, where $s_1$ and $s$ satisfy $0\leq s_1\leq s\ll q$,
$\beta_1,\ldots,\beta_{s_1}$ are the constant coefficients, and
$\beta_{s_1+1}(t),\ldots, \beta_{s}(t)$ are the functional coefficients.
We treat (\ref{modelsemivarying0}) as the final model, and estimate
both the constant and varying coefficient functions with the covariance
structure of $\varepsilon(t)$ taken into account.

To the best of our knowledge, for the present setup, both screening and
simultaneous variable selection and structure identification have not
been studied before, and the estimation methods are new. Note that $p$
is fixed and the structure identification method is a model selection
approach in \cite{ZFS2009}.
We show both theoretically and numerically that, for $p$ of any
exponential order of $n$, based on working independence, the proposed
NIS procedure can keep the relevant variables with high probability.
In addition, we relax the conditions on the threshold parameter as
compared to those in the literature \cite{FFS2011,FMD2013,LLW2013}.
A consequence is that the dimension after the NIS step can be
controlled at a moderate order which fulfills the conditions on the
dimensionality in the subsequent group SCAD step. This provides the
theoretical ground for our new sequential screening and variable
selection approach. In addition, we discuss the computation and tuning
parameter selection issues.

In Section~\ref{secscreening}, our NIS procedure is introduced and
its theoretical properties are studied.
The group SCAD procedure for simultaneous variable selection and
structure identification, and its consistency, sparsity and oracle
properties are given in Section~\ref{secgscad}. The refined
estimation procedure for estimating the constant and varying
coefficients in the final model (\ref{modelsemivarying0}) is detailed
in Section~\ref{secestimation}. Results of a simulation study and
application to the yeast cell cycle data are reported and discussed in
Section~\ref{secnumeric}. Proofs of the theorems and some lemmas are
placed in \hyperref[secproofgs]{Appendix} and the supplementary material \cite{CHLP2014}. %\ref{suppA}.

%%%%%%%%%%%%%%%%%%%%%%%%%%%%%%%%%%%%%%%%%%%%%%
%%%%%%%%%%%%%%%%%%%%%%%%%%%%%%%%%%%%%%%%%%%%%% NIS section
%s2 #&#
\section{Nonparametric independence screening}\label{secscreening}

%%%%%%%%%%%%%%%%%%%%%%%%%%%%%%% Notation
%We introduce some notation first.
Denote the Euclidean norm and the sup norm of a vector $v$ by $|v|$ and
$|v|_\infty$, respectively. Also, for a matrix $A=(a_{ij})$, define
$|A|
= \sup_{|x|=1}|Ax|$ and $|A|_\infty= \sup_{i,j} |a_{ij}|$.
Denote the sup norm and the $L_2$ norm of a
function $g$ on $[0,1]$ by $\| g\|_\infty$ and $\|g \|_{L_2}$,
respectively.
%%%%%%%%%%%%%%%%%%%%%%%%%%%%%%% Definitions of two norms
In order to describe and examine our procedures,
we define, respectively, the empirical and theoretical
inner products of two vector-valued stochastic processes
$ {\mathbf u}(t) \in\mathbb{R}^k$ and~$ {\mathbf v}(t) \in\mathbb{R}^l$~by
\[
\bigl\langle{\mathbf u}, {\mathbf v}^T \bigr\rangle_n =
\frac{1}{n}\sum_{i=1}^n
\frac{1}{m_i} \bigl({\mathbf u}_i(t_{i1}), \ldots, {\mathbf
u}_i(t_{im_i})\bigr) \bigl({\mathbf v}_i(t_{i1}),
\ldots, {\mathbf v}_i(t_{im_i})\bigr)^T
\]
and
\[
\bigl\langle{\mathbf u}, {\mathbf v}^T \bigr
\rangle  = \RE\bigl\{ \bigl\langle{\mathbf u}, {\mathbf v}^T \bigr
\rangle_n \bigr\},
\]
where %${\mathbf u}(t)$ and ${\mathbf v}(t)$ are vector-valued stochastic
%processes on $[0,1]$ and
$\{{\mathbf u}_i(t)\}_{i=1}^n$ and $\{{\mathbf v}_i(t)\}_{i=1}^n$
are i.i.d. samples of ${\mathbf u}(t)$ and ${\mathbf v}(t)$. %Note that $<{
%matrices.
When ${\mathbf u}(t)$ is not stochastic,
we should take ${\mathbf u}_i(t) = {\mathbf u}(t)$. When $k=1$, we define,
respectively, $\| u \|_n$ and $\| u \|$ by
$
\| u \|_n^2=\langle u, u \rangle_n$ and $\| u \|^2=\langle u,
u \rangle$.
Note that for any square integrable function $g$
on $[0,1]$, %we have
$
C_1\| g \|_{L_2} \le\| g \| \le C_2 \| g \|_{L_2}
$
uniformly in $g$.

%s2.1 #&#
\subsection{Nonparametric independence screening algorithm}

Consider the full model (\ref{modelfull}). Define the set of indices
of relevant covariates
by
\[
{\mathcal M}_\kappa= \bigl\{ k\ge1 | \| \beta_k
\|^2 \ge C_{\kappa1} n^{-2\kappa}L \bigr\},
\]
for some positive constant $\kappa$.
Here, $L$ is the dimension of the \mbox{B-}spline basis.
%Also, note that Assumption M2(2) given in Section \ref{theorynis} is
%a kind of sparsity assumption.
Under the sparsity Assumption~\ref{assM2}(2) given in Section~\ref{theorynis},
we can carry out the nonparametric independence screening (NIS) %by
%ranking estimates of $b_k$, $k=1, \ldots, p$,
prescribed in the following.

%Now we describe our screening procedure, which is called nonparametric
%independence screening (NIS). In parametric settings, the independence
%screening procedure was proposed for linear models in \cite{FL2008}
%and studied for generalized linear models in \cite{FS2010}. The NIS
%procedure in \cite{FFS2011} adopts the ideas to the (semiparametric)
%additive regression models. It can deal with the case when $p$ is very
%large compared with the sample size $n$. However, it does not possess
%the consistency property in variable selection.

Similar to (3) of \cite{FMD2013}, we consider for each $k=1,\ldots,p$
a marginal model for $y(t)$ and $x^{( k )}( t )$ defined by
%
%e2.1 #&#
\begin{equation}
\label{modelmarginal} y(t)= a_k(t) + b_k(t)x^{( k )}( t )
+ \eta_k(t),
\end{equation}
where $a_k(t)$ and $b_k(t)$ are given by
$
\mathop{\arg\min}_{a,b\in\mathcal{L}^2[0,1]}\| y- a - b x^{(k)}\|
^2$. Alternatively, \cite{LLW2013} employed a conditional correlation approach.
%%%%%%%%%%%%%%%%%%%%%%%%%%% Definitions of the covariates W
%%%%%%%%%%%%%%%%%%%%%%%%%%%
%Under smoothness conditions specified in Assumption M1, we can
%approximate $b_k$ by using
%an equi-spaced \mbox{B-}spline basis of order 3 on [0,1].
Let\vspace*{1pt} $\bB(t) =
(B_1(t), \ldots, B_L(t))^T$ be an equispaced \mbox{B-}spline basis of order 3
on $[0,1]$, where $L$ is the dimension of the basis. Write
$\bB({\mathbf t}_i)=(\bB(t_{i1}), \ldots, \bB(t_{im_i}))$.
%Note that $\bB(\bt_i)$ is an $L \times m_i$ random matrix.
Then, % the coefficient function
under the smoothness conditions specified in Assumption~\ref{assM1}, $b_k(t)$ in
(\ref{modelmarginal}) can be approximated by some linear
combination\vadjust{\goodbreak}
of $\bB(t)$.
Thus, %based on the sample $(y(\bt_i),\bx(\bt_i)), i=1\ldots,n$,
we can estimate $b_k$ by minimizing the following objective
function: %of $\gamma_1\in\RR^L$ and $\gamma_2\in\RR^L$,
%
%e2.2 #&#
\begin{equation}
\label{estimationmarginal}
\qquad\bigl\| y - \gamma_1^T \bB-
\gamma_2^T x^{(k)}\bB\bigr\|_n^2
\equiv\bigl\| y - \gamma_1^T \bB- \gamma_2^T
\bW_k \bigr\|_n^2, \qquad\gamma_1,
\gamma_2\in\mathbb{R}^L.
\end{equation}
Note that the regressors $\bW_k(t)$ and $\bW_k({\mathbf t})$, and their
sample versions $\bW_{ik}(t)$ and $\bW_{ik}({\mathbf t}_i)$, for the
above \mbox{B-}spline estimation of $b_k$ are given by
%
%e2.3 #&#
\begin{eqnarray}\label{definitionmatrixw}
\bW_k(t)&=& x^{( k )}( t )\bB(t) = \bigl(W_{k1}(t),
\ldots, W_{kL}(t)\bigr)^T \in\mathbb{R}^L,\nonumber
\\
\bW_k({\mathbf t})&=& \bigl(\bW_k(t_1), \ldots,
\bW_k(t_m) \bigr) \in\mathbb{R}^{L\times m},
\nonumber\\[-8pt]\\[-8pt]
\bW_{ik}(t)&=& \bigl(W_{ik1}(t), \ldots, W_{ikL}(t)
\bigr)^T = x_{ i }^{( k )}( t )\bB(t) \in
\mathbb{R}^L,\qquad i=1,\ldots,n,\hspace*{-30pt}
\nonumber
\\
\bW_{ik}({\mathbf t}_i)&=& \bigl(
\bW_{ik}(t_{i1}), \ldots, \bW_{ik}(
t_{im_i}) \bigr) \in\mathbb{R}^{L\times m_i},\qquad i=1,\ldots,n.
\nonumber
\end{eqnarray}
%
%Note that $\bW_k(\bt)$ and $\bW_{ik}(\bt_i)$ are $L\times m$ and $L
Writing $\widehat\gamma_{1k}$ and $\widehat\gamma_{2k}$
for the minimizer of (\ref{estimationmarginal}), we define the
\mbox{B-}spline estimator of~$b_k$ by
%
%e2.4 #&#
\begin{equation}
\label{estimatormarginal} \widehat b_k(t) = \widehat\gamma_{2k}^T
\bB(t).
\end{equation}
Given %the \mbox{B-}spline estimates
$\widehat b_k$, $k=1,\ldots,p$, we carry out the nonparametric
independence screening %by employing $\| \widehat b_k\|_n$
and define the index set of selected covariates, denoted as $\widehat
{\mathcal
M}_\kappa$, by
%
%e2.5 #&#
\begin{equation}
\label{screening} \widehat{\mathcal M}_\kappa= \bigl\{ k \ge1 | \|
\widehat
b_k\|_n^2 \ge C_{\kappa3}n^{-2\kappa}
L\bigr\}
\end{equation}
for some sufficiently small positive constant $C_{\kappa3}$ satisfying
$C_{\kappa3}
< C_{\kappa2}/2$, where $C_{\kappa2}$ is given in Assumption~\ref{assM2}(1).

Intuitively, we may still have too many irrelevant variables kept in
the analysis if the threshold parameter $\kappa$ in (\ref{screening})
is chosen too large. On the other hand, we may run into the danger of
screening out some of the relevant variables if it is chosen too small.
In the literature, the screening step is immediately followed by the
model fitting step, thus some iterative screening procedure is employed
to control the false selection rate \cite{FL2008,FS2010}. We avoid
such time-consuming iterations by adding between these two steps a
variable selection step, given in Section~\ref{methodgroupSCAD}. The
theory given in Section~\ref{theorynis} guarantees that, with proper
choices of $\kappa$ and $L$, by the first screening step we can reduce
the dimensionality to a moderate order with the false negative rate
under control. This allows the next variable selection step to possess
the sparsity and oracle\vspace*{2pt} properties given in Section~\ref{theorygroupSCAD}. In practice, we sort the $ \| \widehat b_k\|_n^2$'s
in the descending order and keep the first $[n^\alpha/\log n]$
variables, for some $2/5 \leq\alpha\leq1$. In the numerical
sections, we took $\alpha=1$.

%s2.2 #&#
\subsection{Theory of the proposed NIS procedure}\label{theorynis}

Here, we collect the technical assumptions on the marginal models given
in (\ref{modelmarginal}).
Let $\bI_d$ denote the identity matrix of
dimension $d$ and let $\# A$ be the number of the elements in a set~$A$.

%%%%%%%%%%%%%%%%%%%%%%%%%%%%%%%% Assumption M1

{\renewcommand{\theass}{M\arabic{ass}}
%as1 #&#
\begin{ass}\label{assM1}
There are positive constants $C_{M0}$ and $C_{M2}$ satisfying
\mbox{(1)--(3)} in the following. For $k=1, \ldots, p$:
\begin{longlist}[(3)]
\item[(1)] $a_k$ and $b_k$ are twice continuously differentiable,\vspace*{1pt}

\item[(2)] $\| a_k\|_\infty\le C_{M0}$ and $\| b_k\|_\infty\le C_{M0}$,\vspace*{1pt}

\item[(3)] $\| a_k''\|_\infty\le C_{M2}$ and $\| b_k''\|_\infty
\le C_{M2}$.
\end{longlist}
\end{ass}

%%%%%%%%%%%%%%%%%%%%%%%%%%%%%%%% Assumption M2

%as2 #&#
\begin{ass}\label{assM2}
For the $\kappa$ in the definition of ${\mathcal M}_\kappa$,
there are positive constants $C_{\kappa1}$ and $C_{\kappa2}$ such that
(1)--(2) in the following hold and we also have (3) given below:
\begin{longlist}[(3)]
\item[(1)] If $\| \beta_k\|^2\ge C_{\kappa1}n^{-2\kappa}L$,
we have $\| b_k\|^2\ge C_{\kappa2}n^{-2\kappa}L$.\vspace*{2pt}

\item[(2)] If $\| \beta_k\|^2 < C_{\kappa1}n^{-2\kappa}L$,
we have $\| \beta_k\|^2= 0$.\vspace*{2pt}

\item[(3)] $n^{1-4\kappa}L/\log n \to\infty$,
$n^{-2\kappa}L =o(1)$, and $ L^{-3} = o ( n^{ -2\kappa}) $.
\end{longlist}
\end{ass}}

Assumption~\ref{assM1} is necessary in order to bound the approximation error
to $a_k$~and~$b_k$ by \mbox{B-}spline bases.
Assumption~\ref{assM2} %below is a counterpart of Assumption C in
requires that the marginal models~(\ref{modelmarginal}) still reflect
the significance
of relevant covariates, and similar assumptions are assumed in the
NIS literature \cite{FFS2011}.
We mention that, in Assumption~\ref{assM2}(2), we require that $\|\beta_k\|^2=0$
merely\vspace*{2pt} for simplicity of presentation, and it is sufficient to replace
it with $\| \beta_k\|^2= o(n^{-2\kappa}L)$. %Theorem \ref{teothm1}
%implies that all the relevant covariates will be selected with high
%probability, due to Assumption M2(1) and the uniform consistency.
%Furthermore, we emphasize that, by exploiting the band diagonal
%property of $\nip{\bB,\bB^T}$, $\ip{\bB,\bB^T}$, $\nip{\bW_k, \bW_l}$,
%$\ip{\bW_k, \bW_l}$, and so on, in Assumption M2(3) we have relaxed
%the conditions on $\kappa$ as compared to those in \cite{FFS2011}.

%%%%%%%%%%%%%%%%%%%%%%%%%%%%%%%%%%%%%%%%%%%% Assumption T

{\renewcommand{\theass}{T}
%as3 #&#
\begin{ass}\label{assT}
(1) For some positive constant $C_{T1}$, we have
$C_{T1} \le\RE[ \{ \widetilde x^{(k)}(t) \}^2]$ for
$t\in[0,1]$ and $k=1,\ldots, p$, where
$\widetilde x^{(k)}(t)= x^{( k )}( t )- \RE\{ x^{( k )}( t )\}$.

(2) For any positive constant $C_1$, there is a positive constant
$C_2$ such that
\[
\RE\bigl[ \exp\bigl\{ C_1 \bigl|\varepsilon_i ({\mathbf
t}_i )\bigr|/\sqrt{m_i} \bigr\} | {\mathbf x}_i({
\mathbf t}_i), {\mathbf t}_i \bigr] < C_2.
\]

(3) Let ${\mathbf x}_{\mathcal M}(t)$ be the covariate vector consisting
of $1$ and all the covariates in~${\mathcal M}_\kappa$. Then there
is a positive constant $C_{T2}$ such that %for any $t\in[0,1]$,
\[
\RE\bigl\{ {\mathbf x}_{\mathcal M}(t) {\mathbf x}_{\mathcal M}(t)^T
\bigr\} \ge C_{T2} \bI_{\#
{\mathcal M}_\kappa+ 1}\qquad\mbox{for any } t\in[0,1].
\]

(4) For some positive constant $C_{T3}$,
${\sup_{t\in[0,1]}|x^{( k )}( t )|\le C_{T3}}$,
$k=1,\ldots, p$.\vspace*{1pt}

(5) For some positive constant $C_{T4}$,
${\sum_{k\in{\mathcal A}}
\sup_{t\in[0,1]}|\beta_k(t)| \le C_{T4}}$,
where ${\mathcal A}= \{ k | 0\le k \le p$ and $\sup_{ t\in[0,1] }
|\beta_k(t)| >0 \} $.\vspace*{1pt}

(6) The functions $\beta_k(t)$, $k=0,\ldots,p$, are twice continuously
differentiable. Besides, $\sum_{k\in{\mathcal A}}
\sup_{t\in[0,1]}|\beta''_k(t)| \le C_{T5}$ for some positive
constant $C_{T5}$.
\end{ass}}

Assumption~\ref{assT}(1)~and~(4) imply that for some
positive constants $C_{1}$ and $C_{2}$, we have
%
%e2.6 #&#
\begin{equation}
\qquad C_{1} \bI_2 \le\RE\pmatrix{ 1 & x^{( k )}( t )
\vspace*{3pt}\cr
x^{( k )}( t ) & \bigl\{x^{( k )}( t )\bigr\}^2}\le
C_{2} \bI_2,\qquad t\in[0,1]\mbox{ and } 1\le k \le p.
\label{eqnh1311}
\end{equation}
Assumption~\ref{assT}(2), (4) and~(5) are technical assumptions needed in order
to apply the exponential inequalities.\vspace*{1pt} Note that $\# {\mathcal A}$ may
increase, but we assume that the signal $\sum_{k\in{\mathcal A}}
x^{(k)}(t)\beta_k(t)$ should not diverge by
imposing Assumption~\ref{assT}(5)~and~(6). Similar conditions are made in
Assumption~D of \cite{FFS2011}.
We can relax T(4) and T(5) slightly, for example,
we can replace $C_{T3}$ and $C_{T4}$ with $C_{T3}\log n$ and
$C_{T4}\log n$
at the expense of multiplying
the present convergence rate by $(\log n)^c$ for some positive $c$. We
can also
relax Assumption~\ref{assT}(6) similarly with conformable changes made in the
approximation error of the \mbox{B-}spline basis.
We need Assumption~\ref{assT}(1) and (\ref{eqnh1311})
for identifiability and estimation of
the marginal models. Assumption~\ref{assT}(3)
is the identifiability condition of the coefficient functions in
model (\ref{eqnh101}). We use Assumption \ref{assT}(5) and (6) to evaluate the
approximation error of the \mbox{B-}spline basis when we consider the
group SCAD variable selection discussed in Section~\ref{secgscad}.

%%%%%%%%%%%%%%%%%%%%%%%%%%%%%%%%%%%% Theorem 1
%
%th2.1 #&#
\begin{teo}
\label{teothm1}
Suppose Assumptions~\ref{assM1},~\ref{assM2}, and~\ref{assT}(1)--(5) hold.
Then
\[
\RP( {\mathcal M}_\kappa\subset\widehat{\mathcal M}_\kappa)
\ge1 - C_{p1}pL \exp\bigl( -C_{p2}n^{1-4\kappa} L\bigr),
\]
where $C_{p1}$ and $C_{p2}$ are positive constants
depending on $C_{\kappa j }$, $j=1,2,3$, and the constants
specified in the above mentioned assumptions.
\end{teo}

Theorem \ref{teothm1} implies that all the relevant covariates will
be selected with high probability, due to Assumption~\ref{assM2}(1) and the
uniform consistency.
Specifically, when $p = O(n^{c_p})$ for any positive $c_p$ we have
$\RP( {\mathcal M}_\kappa\subset\widehat{\mathcal M}_\kappa)
\to1$, if $\kappa$ satisfies Assumption~\ref{assM2}(3). Under the smoothness
Assumption~\ref{assM1}(1),
the optimal rate of $L$ is $L=c_Ln^{1/5}$
for some\vspace*{1pt} positive $c_L$. In this case, Assumption~\ref{assM2}(3) reduces to
$
n^{6/5-4\kappa}/\log n \to\infty,
n^{1/5-2\kappa}=o(1)$, and
$n^{2\kappa-3/5}=o(1)$.
Then a sufficient condition on $\kappa$ is that
%
%e2.7 #&#
\begin{equation}
\label{kappa} 1/10 < \kappa<3/10.
\end{equation}
Thus, the proposed screening procedure may reduce the number of
covariates drastically. However, $\# \widehat{\mathcal M}_\kappa$
may be still too large to apply any variable selection
procedures with consistency property. Fortunately, we have succeeded
in giving an upper bound on $\# \widehat{\mathcal M}_\kappa$, as
given in
Theorem \ref{teothm2}, which circumvents such situations. We
emphasize that condition (\ref{kappa}) is weaker than those in the
literature: Theorem~1 of~\cite{FMD2013} requires that $n^{1-4\kappa}
L^{-3}\to\infty$ which reduces to $ \kappa< 1/10$ when $L$ is of the
order $n^{1/5}$, and in Theorem~2 of \cite{LLW2013} the condition on
$\kappa$ implies $ \kappa< 1/10$ as well. This improvement is crucial
for us to obtain a tighter upper bound in Theorem~\ref{teothm2}, as
compared to that in \cite{FMD2013} (no upper bound is provided in
\cite{LLW2013}), which leads to~(\ref{numM}). We succeed in achieving this improvement by exploiting
the band diagonal property of $\langle\bB,\bB^T \rangle_n$,
$\langle\bB,\bB^T \rangle$, $\langle\bW_k, \bW_l \rangle_n$,
$\langle\bW_k, \bW_l \rangle$, and so on. %, in Assumption M2(3) we
%have relaxed the conditions on $\kappa$ as compared to those in

In order to state Theorem \ref{teothm2}, we need a little more notation.
Define %$\overline\bW_k$ and $\overline\bW$ by
%
%e2.8 #&#
\begin{eqnarray}
\overline\bW_k & =& \bW_k -A_k\bB\quad\mbox{and}\quad% A_k=\ip{\bW_k,\bB^T}\ip{\bB, \bB^T}^{-1},
\overline\bW= \bigl(\overline
\bW_1^T, \ldots, \overline\bW_p^T
\bigr)^T,\label{eqnh303}
\end{eqnarray}
where $ A_k=\langle\bW_k,\bB^T \rangle\langle\bB, \bB^T \rangle
^{-1}$. Note that $\overline\bW_k$ %is an $L$-dimensional stochastic
%process on $[0,1]$
and $\overline\bW$ are, respectively, $L$- and $pL$-dimensional stochastic
processes on $[0,1]$. Besides we define
%
%e2.9 #&#
\begin{equation}
\overline\Sigma= \bigl\langle\overline\bW, \overline\bW^T \bigr
\rangle, \label{eqnh305}
\end{equation}
which is a $pL\times pL$
matrix. We write $\lambda_{\max}(A)$ and
$\lambda_{\min}(A)$ for the maximum and minimum
eigenvalues of a symmetric matrix $A$, respectively.

%%%%%%%%%%%%%%%%%%%%%%%%%%%%%%%%%%%%%%%% Theorem 2
%
%th2.2 #&#
\begin{teo}
\label{teothm2}
Under the same assumptions as in Theorem \ref{teothm1} except Assumption~\ref{assM2}(2),
we have for some positive constant $C_{\kappa4}$,
\[
\RP\bigl( \# \widehat{\mathcal M}_\kappa\le C_{\kappa4}
n^{2\kappa} \lambda_{\max}( \overline\Sigma) \bigr) \ge1 -
C_{p1}pL \exp\bigl( -C_{p2}n^{1-4\kappa} L \bigr),
\]
where $C_{p1}$ and $C_{p2}$ are the same constants as in Theorem \ref{teothm1}.
\end{teo}

Theorem \ref{teothm2} implies that, with high probability, the number of
variables selected by our screening procedure will not be large. %, due
%to the upper limit.
Note that it does not require Assumption~\ref{assM2}(2). This means that, %we do
%not have to care about the size of $ \|b_k\| $ when $ \|\beta_k\| $ is
%small (0 here) since relevant variables will be selected and the
%number of selected ones is not large. In other words,
although some of the irrelevant covariates (with $ \|\beta_k\| $
small) may be included in $\widehat{\mathcal M}_\kappa$ merely
because they have large values of $ \|b_k\| $, the number of such
variables is limited. Furthermore,
they will be removed by the subsequent variable selection procedure
given in Section~\ref{secgscad}.

Define $\widetilde\bW_k(t)$, $\widetilde\bW_k({\mathbf t})$,
$\widetilde\bW_{ik}(t)$ and $\widetilde\bW_{ik}({\mathbf t}_i)$
by replacing $x^{( k )}( t )$ and $x_{ i }^{( k )}( t )$ in the
definitions of $\bW_k(t)$, $\bW_k({\mathbf t})$,
$\bW_{ik}(t)$ and $\bW_{ik}({\mathbf t}_i)$ given in (\ref{definitionmatrixw}) with
\[
\widetilde x^{(k)}(t)= x^{( k )}( t )- \RE\bigl\{
x^{( k )}( t )\bigr\}\quad\mbox{and}\quad \widetilde x_i^{(k)}(t)=
x_{ i }^{( k )}( t )- \RE\bigl\{ x_{ i }^{( k )}(
t )\bigr\},
\]
respectively. It is easy to see, by properties of orthogonal
projection that
\[
\bigl\langle\widetilde\bW, \widetilde\bW^T \bigr\rangle\le\overline
\Sigma\le\bigl\langle\bW, \bW^T \bigr\rangle,
\]
where $\widetilde\bW= ( \widetilde\bW_1^T,
\ldots, \widetilde\bW_p^T)^T$ and
$\bW= ( \bW_1^T, \ldots, \bW_p^T)^T$. The maximum
eigenvalue of $ \langle\bW, \bW^T \rangle$ may tend to infinity
very quickly with $p$. However,\vspace*{1pt} since we do a kind of centerization to
$\bW$ and obtain $\overline\bW$ as in (\ref{eqnh303}), we
conjecture that $ \overline
\Sigma$ is very close to $\langle\widetilde\bW, \widetilde\bW^T
\rangle$ under
some regularity conditions. If the maximum eigenvalues of the two
matrices have the same order, and for some positive $K_n$,
$
\lambda_{\max} ( \RE\{ \widetilde{\mathbf x}(t) \widetilde{\mathbf
x}(t)^T \} )
\le K_n
$
uniformly in $t$, then %we have
\[
\lambda_{\max} ( \overline\Sigma) \le C_1
L^{-1}K_n \quad\mbox{and}\quad\# \widehat M_\kappa
\le C_2 n^{2\kappa}L^{-1}K_n
\]
with probability tending to 1. %, {\color{red} as in \cite{FFS2011}}.
Suppose $L$ is chosen to be of the optimal order $n^{1/5}$. Then, for
$\kappa$ satisfying condition (\ref{kappa}), this implies that
%
%e2.10 #&#
\begin{equation}
\label{numM} \# \widehat M = O_p\bigl(n^{2/5-\eta}\bigr)
K_n \qquad\mbox{for some } 0< \eta< 2/5.
\end{equation}
Thus, when $K_n$ is bounded, $\# \widehat M$ fulfills the requirement
(\ref{eqnq}) on $q$ in the subsequent variable selection step given
in Section~\ref{secgscad}. In addition, if we choose a smaller value
of $\kappa$, we can further allow a moderately increasing $K_n$.

%%%%%%%%%%%%%%%%%%%%%%%%%%%%%%% gscad
%s3 #&#
\section{Variable selection and structure identification}\label{secgscad}

We can remove a lot of irrelevant covariates by the NIS procedure given
in Section~\ref{secscreening}. However, it does not have
the consistency property in selecting the important variables. In this section,
%considering the case that the number of variables is below some order
%of $n$,
we propose a group SCAD estimator for variable selection and structure
identification, and establish
its consistency, sparsity and oracle properties. Here, we denote the
number of covariates by $q$, instead of $p$ as in Section~\ref{secscreening}.
This distinction is necessary. When the dimensionality $p$ is very
large, we have to employ some screening procedure %, such as our NIS
%procedure,
before we can carry out any variable selection procedure. In this case,
$p$ and $q$ are, respectively, the number of variables before and after
the screening procedure is applied. For simplicity of notation, we
still denote as $x^{(1)}, \ldots, x^{(q)}$ the variables selected by
the NIS algorithm. When the $p$ is not very large, % i.e. $p$ satisfies
%the conditions on $q$ given in this section,
we can simply take $q=p$ and proceed directly with the group SCAD
procedure.% given in Section \ref{methodgroupSCAD}.

%s3.1 #&#
\subsection{Group SCAD procedure}\label{methodgroupSCAD}

Suppose we are given $y_i({\mathbf t}_i), x_i^{(k)}({\mathbf t}_i)$,
$i=1,\ldots,n$, $k=1,\ldots,q$ and
consider the varying coefficient model (\ref{modelnis0}).
%y_i(\bt_i) = \beta_0 (\bt_i) + \sum_{k=1}^q \xb{i}{k}{\bt_i}\beta_k(
%where $\varepsilon_i(\bt_i), i=1,\ldots,n$, are the same as in model
%(\ref{modelfull}).
To estimate the coefficient functions $\beta_k(t), k=0,1,\ldots,q$, first we define %$\bgamma= (\gamma_0^T, \ldots,\gamma_q^T)^T \in
%
%e3.1 #&#
\begin{equation}
l_q(\bgamma)= \Biggl\| y- \gamma_0^T\bB- \sum
_{k=1}^q \gamma_k^T
\bW_k \Biggr\|_n^2, % \mbox{where } \bgamma= (\gamma_0^T, \ldots,
\label{eqnh403}
\end{equation}
where $\bgamma= (\gamma_0^T, \ldots,
\gamma_q^T)^T \in\mathbb{R}^{(q+1)L}$. When $q$ is fixed and
sufficiently small, based on working independence, % of the error
%process,
we can estimate $\beta_k(t)$ %the coefficients in model (
by minimizing the objective function $l_q(\bgamma)$. Denoting the
minimizer by
$\widetilde\bgamma= (\widetilde\gamma_0^T,
\ldots, \widetilde\gamma_q^T)^T$, for $k=0,1,\ldots,q$, we can
estimate $\beta_k(t)$ by
%
%e3.2 #&#
\begin{equation}
\widetilde\beta_k (t) = \widetilde\gamma_k^T
\bB(t). \label{eqnh405}
\end{equation}
Recall that $L$ is the dimension of the \mbox{B-}spline basis. Suppose $q$ satisfies
%
%e3.3 #&#
\begin{equation}
q= o \bigl(\sqrt{n/(L\log n)} \bigr). \label{eqnh401}
\end{equation}
This restriction is necessary since the Hessian
matrix of the objective function $l_q(\bgamma)$ given in (\ref{eqnh403})
must be positive definite.
Note that \cite{WLH2008} imposed similar conditions in the case where
$q$ is a fixed constant.
%Recall that $L$ is the dimension of the \mbox{B-}spline basis. %and tends to $
When $L$ is taken as the optimal order $n^{1/5}$, condition (\ref{eqnh401})
reduces to
%
%e3.4 #&#
\begin{equation}
\label{eqnq} q=o \bigl(n^{2/5}(\log n)^{-1/2}\bigr).
\end{equation}
%
%We repeat that, in Section \ref{secgscad}, Appendix

When $q$ is relatively large and a lot of the covariates
seem to be irrelevant, we would add a penalty term to $l_q(\bgamma)$
given in (\ref{eqnh403}),
such as the group SCAD or the adaptive group Lasso penalty, and then conduct
variable selection and estimation simultaneously. After the variable
selection step, if necessary,
we can estimate the coefficient functions of the selected variables
again without the penalty term.
Besides, we are also interested in structure identification. That is,
some of the coefficient functions %$\beta_k$ associated with the
%relevant variables
may be constant while the others are time-varying.
%In this case, the model is further reduced to a semivarying
%coefficient model.
As mentioned in Section~\ref{secsetup}, when there is no a priori
knowledge on
which of the coefficient functions are indeed constant, treating the
constant coefficients as time-varying would
result in a loss in the convergence rate. Thus, an important issue is
to identify them based on data. To this end, we can add another penalty
term to regularize the estimated coefficient functions. A similar kind
of penalty term was used in \cite{YH2012} for Cox proportional hazard
models with time-varying coefficients.

Now we define our group SCAD penalty term for simultaneous variable
selection and structure identification. %Write $g_k(t)= \gamma_k^T \bB
%(t)$, $k=0,1,\ldots,q$.
First, we introduce an orthogonal decomposition of $g_k(t)
\equiv\gamma_k^T \bB(t)$ with respect to the $L_2$ norm by
%
%e3.5 #&#
\begin{equation}
g_k(t) = (g_{k})_c + (g_{k})_f(t),
\label{eqnh411}
\end{equation}
where
$
(g_{k})_c=\int_0^1 g_k(t)\,dt$
and $(g_{k})_f(t) = g_k (t) - (g_{k})_c$.
Then, we have
$
\| g_{k}\|_{L_2}^2 = |(g_{k})_c|^2 + \|
(g_{k})_f \|_{L_2}^2$.
%%%%%%%%%%%%%%%%%%%%%%%%%%%%%%%%%%% gSCAD penalty
%In addition, clearly $|(g_{k})_c|= \|(g_{k})_f \|_{L_2}=0$ when $g_k$
%is the zero function and $ \|(g_{k})_f \|_{L_2}=0$ when $g_k$ is a
%non-zero constant.
Let $p_\lambda(
\cdot) $ be the SCAD function given by
%
%e3.6 #&#
\begin{equation}
\label{eqnh409} \qquad p_\lambda(u ) = \cases{ \lambda u, &\quad if $0 \le u
\le
\lambda$,
\vspace*{3pt}\cr
-\bigl(u^2-2a_0\lambda u +
\lambda^2 \bigr)/\bigl\{2(a_0 -1 )\bigr\}, &\quad if $
\lambda< u \le a_0 \lambda$,
\vspace*{3pt}\cr
(a_0+1)
\lambda^2/2, &\quad if $u > a_0\lambda$,}
\end{equation}
where $a_0$ is a constant larger than 1. We take $a_0 = 3.7$ as
suggested by \cite{FL2001}. Our
group SCAD penalty is defined by
$
\sum_{k=1}^q \{ p_{\lambda_1}(|(g_{k})_c|) +
p_{\lambda_2}(\|(g_{k})_f\|_{L_2}) \}$,
where $g_k= \gamma_k^T \bB$, $k=0,\ldots, q$.
We specify the values of $\lambda_1$ and $\lambda_2$ later.
Our objective function for our group SCAD estimator
is then given by
%%%%%%%%%%%%%%%%%%%%%%%%%%%%%%%%%%%%%%%%%%%%%%%%% Q_q
%
%e3.7 #&#
\begin{equation}
Q_q(\bgamma) = l_q( \bgamma) + \sum
_{k=1}^q \bigl\{ p_{\lambda_1}
\bigl(\bigl|(g_{k})_c\bigr|\bigr) + p_{\lambda_2}\bigl(
\bigl\|(g_{k})_f \bigr\|_{L_2}\bigr) \bigr\}.
\label{eqnh415}
\end{equation}
Based on $Q_q(\bgamma)$, we can carry out variable selection,
structure identification, and estimation simultaneously
by the following procedure:
%
%e3.8 #&#
\begin{equation}
\label{estimatescad} \qquad\widehat\bgamma=\bigl( \widehat\gamma_0^T,
\ldots, \widehat\gamma_q^T\bigr)^T = \mathop{
\arg\min}_{\bgamma\in
\mathbb{R}^{(q+1)L}} Q_q( \bgamma)\quad\mbox{and}\quad \widehat
\beta_k= \widehat\gamma_k^T\bB, k=0,\ldots, q.
\end{equation}
Then we can choose the significant covariates as those $x^{(k)}$ with
$\|\widehat\beta_k\|_{L_2}>0$ and identify the constant coefficients
by the criterion $\|(\widehat\beta_k)_f\|_{L_2}=0$. We call $\widehat
\beta_k$ the group SCAD estimator.

To compute the group SCAD estimator given in (\ref{estimatescad}), we
use the approximation to the SCAD function suggested in \cite{FL2001}:
$
p_\lambda(u) \approx p_\lambda(u_0)+\frac{1}{2}(p'_{\lambda
}(u_0)/\break u_0)(u^2-u_0^2) $,
for $u$ in a neighborhood of any given $u_0\in\mathbb{R}^+$.
Define
$
\tau_j=\tau^{-1}\int_0^1 B_j(t) \,dt $,
and
$\overline B_j = \sqrt{L} (
B_j - \tau_1^{-1}{\tau_j} B_1), j=0,1,\ldots,L$,
where\vspace*{1pt} $\{B_0(t),\break  B_1(t), \ldots, B_L(t)\}$ is the \mbox{B-}spline basis on $[0,1]$.
Then it will be convenient to use the new basis $(1,\overline B_2,
\ldots, \overline B_L)$ when we calculate the group SCAD penalty term.
The number of covariates $q$ after the NIS step should be small enough
to calculate the least squares estimates, which can be used as the
initial estimates in the iterative algorithm for finding $\widehat
\bgamma$. To select the tuning parameters $\lambda_1$ and $\lambda
_2$ in~(\ref{eqnh415}), we treat them as equal $\lambda_1=\lambda
_2=\lambda$ and use\vspace*{1pt} the BIC criterion to select $\lambda$:
$
\mbox{BIC}(\lambda) %= \Big\| y- \widehat\gamma_0(\lambda)^T\bB-
%K} \log N
= \| y- \widehat\beta_0(\lambda)- \sum_{k=1}^q \widehat\beta
_k(\lambda) x^{(k)} \|_n^2 + {\mathcal K} \log{\mathcal N}$,
where\vspace*{1pt} $\widehat{\beta}_k(\lambda)$ is the group SCAD estimate based
on~$\lambda$, ${\mathcal K}$ is the number of parameters in the fitted
model, and ${\mathcal N}=\sum_{i=1}^{n} m_i$.
A similar BIC criterion was suggested by \cite{WLT2007}, and a
generalized information criterion was considered by \cite{FT2013} for
tuning parameter selection in penalized likelihood models.

%s3.2 #&#
\subsection{Asymptotic properties of the group SCAD procedure}\label{theorygroupSCAD}

In this section, we state the consistency, sparsity, and oracle
properties of the proposed group SCAD estimator given in (\ref{estimatescad}).
The proofs of the theorems are deferred to \hyperref[secproofgs]{Appendix}.
First, we state the sparsity assumption.
%We define the convergence rate $r_{qn}$ in Assumption S(2) later in
%this section.
We can relax Assumption~\ref{assS}(2) in some sense.
See Remark \ref{remarkrmk1} %at the end of this subsection
for more details.

%%%%%%%%%%%%%%%%%%%%%%%%%%%%%%%%%%%% Assumption S

{\renewcommand{\theass}{S}
%as4 #&#
\begin{ass}\label{assS}
(1) There is a positive integer $s<q$ such that the following hold:
\begin{description}
\item[] for $k=1,\ldots,s$, $|(\beta_{k})_c|/\lambda_1\to\infty$ if
$|(\beta_{k})_c|>0$ and $\|(\beta_{k})_f\|_{L_2}/\lambda_2\to
\infty$ if $\|(\beta_{k})_f\|_{L_2}>0$;
for $k=s+1,\ldots,q$, $\|\beta_{k}\|_{L_2}=0$;
the above divergence is uniform in $k=0,1,\ldots, s$.
\end{description}

(2) $\lambda_1/r_{qn}\to\infty$ and $\lambda_2
/r_{qn}\to\infty$, where $r_{qn}$ is defined in (\ref{eqnh503}). %,
%is the convergence rate of the group SCAD estimator.
\end{ass}}

%It is crucial to define suitable function spaces and norms on the
%function spaces when we consider theoretical properties of spline
%estimators.
We define the spline estimation space, denoted as $\bG$,
by
\[
\bG= \bigl\{ {\mathbf g}= (g_0, \ldots, g_q)^T
| g_k=\gamma_k^T \bB, k=0,1,\ldots, q \bigr\}
\]
and $\bG_0$, which we may call the oracle space under Assumption~\ref{assS}, by
\begin{eqnarray*}
\bG_0 &= & \bigl\{ {\mathbf g}\in\bG| (g_k)_c=0\mbox{ if } \bigl|(\beta_k)_c\bigr|=0
\mbox{ and }(g_k)_f=0\mbox{ if } \bigl\|(
\beta_k)_f\bigr\|_{L_2}=0,
\\
&&\hspace*{227pt}  k=1,\ldots,q \bigr\}.
\end{eqnarray*}
%
%%%%%%%%%%%%%%%%%%%%%%%%%%%%%%%%%%% L_2 and L_\ifty norms
We introduce two norms on $\bG$ here. % to establish our main results.
For ${\mathbf g}= (g_0,g_1, \ldots, g_q)^T \in\bG$, define
$
\| {\mathbf g}\|_{L_2}^2= \sum_{k=0}^q \| g_k \|_{L_2}^2$ and $\| {\mathbf
g}\|_\infty= \sum_{k=0}^q \| g_k \|_\infty$.
%%%%%%%%%%%%%%%%%%%%%%%%%%%%%%% Approximation error \rho_{qn}
The approximation error of spline functions to $\bbeta=(\beta_0,
\ldots, \beta_q)^T$, denoted as $\rho_{qn}$, affects the convergence
rates of the least squares and the group SCAD estimators, and we define
it by
$
\rho_{qn} = \sup_{\bbeta}\inf_{{\mathbf g}\in\bG}\| \bbeta- {\mathbf g}
\|_\infty$,
where the supremum is taken over $\bbeta$ satisfying\vspace*{1pt} Assumption~\ref{assT}(5)--(6).
Corollary 6.26\vspace*{1pt} of \cite{Schumaker2007} and Assumption~\ref{assT}(5)--(6) imply
that $\rho_{qn} \le C_\rho L^{-2}$ for some positive constant
$C_\rho$.
%%%%%%%%%%%%%%%%%%%%%%%%%%%%%%%%%%%%%% Convergence rate r_{qn}
Before we state Theorems \ref{teothm3}--\ref{teothm5},
we define the convergence rates of the least squares and the group SCAD
estimators, respectively, denoted as $r_{qn}$ and $r_{sn}$, by
%
%e3.9 #&#
\begin{equation}
r_{qn}=\max\bigl\{ (qL/n)^{1/2}, \rho_{qn}\bigr\}
\quad\mbox{and}\quad r_{sn}=\max\bigl\{ (sL/n)^{1/2},
\rho_{qn}\bigr\}, \label{eqnh503}
\end{equation}
where $s$, defined in Assumption~\ref{assS}, is the number of relevant
variables.\vadjust{\goodbreak}

We state two technical assumptions here. Set
%
%e3.10 #&#
\begin{equation}
\Sigma_n = \bigl\langle\bigl(\bB^T \town^T
\bigr)^T, \bigl(\bB^T \town^T\bigr) \bigr
\rangle_n\quad\mbox{and}\quad \Sigma= \RE\{ \Sigma_n \},
\label{eqnh406}
\end{equation}
where $\town= (\bW_1^T, \ldots, \bW_q^T)^T$.
A sufficient condition for Assumption~\ref{assV} is
$
\lambda_{\min} \RE\{ {\mathbf x}(t){\mathbf x}(t)^T \} \ge C
$
uniformly in $t$ for some positive $C$.

%%%%%%%%%%%%%%%%%%%%%%%%%%%%%%%%%%%%%%%%%%% Assumption E

{\renewcommand{\theass}{E}
%as5 #&#
\begin{ass}\label{assE}
There is a positive constant $C_{E}$
such that uniformly in $t$,
$
\RE\{ \varepsilon_i({\mathbf t}_i ) \varepsilon_i
({\mathbf t}_i )^T | {\mathbf x}_i ({\mathbf t}_i), {\mathbf t}_i \}
\le C_E \bI_{m_i}$.
\end{ass}}

{\renewcommand{\theass}{V}
%as6 #&#
\begin{ass}\label{assV}
There is a positive constant $C_{V}$ such that
$
\lambda_{\min}( \Sigma)\ge C_{V}/L.
$

%%%%%%%%%%%%%%%%%%%%%%%%%%%%%%%%% Main theorems
%In this section,

In Theorem \ref{teothm4},
we derive the $L_2$ convergence rate of the group SCAD estimator %$(
given in (\ref{estimatescad}). %, and we establish some of its
%desirable properties in Theorem \ref{teothm5}, under the sparsity
%assumption (Assumption S).
Before that, in Theorem \ref{teothm3} we deal with
the $L_2$ convergence of the \mbox{B-}spline
estimator %$(\widetilde\beta_0, \ldots, \widetilde\beta_q)^T =(
given in (\ref{eqnh405}).
\end{ass}}
%where

%The $L_2$ convergence rate of the \mbox{B-}spline estimator given in (

%%%%%%%%%%%%%%%%%%%%%%%%%%%%%%%%%%%%%% Theorem 3
%
%th3.1 #&#
\begin{teo}
\label{teothm3}
Suppose that Assumptions~\ref{assT}(4)--(6), \ref{assV} and~\ref{assE} hold. Then
\[
\bigl\| \bigl(\widetilde\gamma_0^T\bB, \ldots, \widetilde
\gamma_q^T\bB\bigr)^T - (\beta_0,
\ldots, \beta_q)^T \bigr\|_{L_2}^2 = \sum
_{k=0}^q \bigl\| \widetilde
\gamma_k^T\bB-\beta_k\bigr\|_{L_2}^2
= O_p \bigl(r_{qn}^2\bigr).
\]
\end{teo}

%%%%%%%%%%%%%%%%%%%%%%%%%%%%%%%%%%%%%% Theorem 4
%
%th3.2 #&#
\begin{teo}
\label{teothm4}
Suppose that Assumptions~\ref{assT}(4)--(6), \ref{assV}, \ref{assE}, and~\ref{assS} hold. Then
with probability tending to 1, there exists a local
minimizer of $Q_q(\bgamma)$ on $\mathbb{R}^{(q+1)L}$, denoted by
$\widehat\bgamma= (\widehat\gamma_0^T,
\ldots, \widehat\gamma_q^T)^T$, such that
\[
\bigl\| \bigl(\widehat\gamma_0^T\bB, \ldots, \widehat
\gamma_q^T\bB\bigr)^T - (\beta_0,
\ldots, \beta_q)^T \bigr\|_{L_2}^2 = \sum
_{k=0}^q \bigl\| \widehat\gamma_k^T
\bB-\beta_k\bigr\|_{L_2}^2 = O_p
\bigl(r_{qn}^2\bigr).
\]
\end{teo}

%In Theorem \ref{teothm5} we state the desirable sparsity and oracle
%properties of the
%local minimizer of $Q_q( \bgamma)$ on $\RR^{(q+1)L}$ in Theorem
Next, we define the sparsity and the oracle properties of estimators.
%under Assumption S(1).

\begin{SP*}
Suppose that Assumption~\ref{assS}(1) holds. Then if an estimator
$\widehat{\mathbf g}=(\widehat g_0, \ldots, \widehat g_q)^T$
of $ (\beta_0,\ldots, \beta_q)^T$ satisfies the conditions below with
probability tending to 1, we say that $\widehat{\mathbf g}$ has the
sparsity property.
\begin{longlist}[(2)]
\item[(1)] For $k=0,\ldots,s$:  $|(\widehat g_{k})_c|>0$ if and only
if $|(\beta_{k})_c|>0$, and $\|(\widehat g_{k})_f\|_{L_2}>0$
if and only if $\|(\beta_{k})_f\|_{L_2}>0$.
\item[(2)] For $k=s+1,\ldots,p$: $\|(\widehat g_{k})_f\|_{L_2}=0$.
\end{longlist}
\end{SP*}

\begin{OP*}
If we\vspace*{1pt} knew the value of %the number of
%important variables
$s$ in Assumption~\ref{assS}(1), we would use the knowledge
and minimize $l_q (\bgamma) $ on the subspace of
$\mathbb{R}^{(q+1)L}$ corresponding to $\bG_0$.
We call this imaginary estimator the oracle estimator. We say that an
estimator has the oracle
property if it is asymptotically equivalent to this
oracle estimator.
\end{OP*}

Theorem \ref{teothm5} is about the sparsity property
and the oracle property of the group SCAD estimator defined in (\ref{estimatescad}).
Note that the existence of the local solution in Theorem \ref{teothm5}
is established in Theorem \ref{teothm4}.

%%%%%%%%%%%%%%%%%%%%%%%%%%%%%%%%%%%%%% Theorem 5
%
%th3.3 #&#
\begin{teo}
\label{teothm5}
Suppose that Assumptions~\ref{assT}(4)--(6), \ref{assV}, \ref{assE}, and \ref{assS} hold.
Let $\{ \eta_n \}$ be a sequence of positive numbers satisfying
$\eta_n \to\infty$, $\lambda_{1}/(\eta_nr_{qn})\to\infty$, and
$\lambda_{2}/(\eta_nr_{qn})\to\infty$.
Then,\vspace*{1pt} with probability tending to 1,
any local minimizer $\widehat\bgamma= (\widehat\gamma_0^T,\ldots, \widehat\gamma_q^T)^T$ of $Q_q( \bgamma)$ satisfying
$\| (\widehat\gamma_0^T\bB, \ldots, \widehat\gamma_q^T\bB)^T
- (\beta_0,\ldots,\break  \beta_q)^T \|_{L_2}\le\eta_nr_{qn}$
is equal to the oracle estimator. We also have
\[
\bigl(\widehat\gamma_0^T\bB, \ldots, \widehat
\gamma_q^T\bB\bigr)^T \in\bG_0
\]
and
\[
\sum_{k=0}^s \bigl\|
\widehat\gamma_k^T \bB-\beta_k
\bigr\|_{L_2}^2= O_p \bigl(r_{sn}^2
\bigr).
\]
\end{teo}

Since $Q_q(\gamma)$ may not be concave,
there may be another local minimizer of $Q_q(\bgamma)$
outside $\{
\bgamma\in\mathbb{R}^{(q+1)L}
|
\| (\gamma_0^T\bB, \ldots, \gamma_q^T\bB)^T
- (\beta_0,\ldots, \beta_q)^T \|_{L_2}\le\eta_nr_{qn}
\}$.
%Remark \ref{remarkrmk1} below is related to the tuning parameter
%selection. %Its proof is outlined in Section \ref{proofrmk1} of the
%online supplementary material.

%%%%%%%%%%%%%%%%%%%%%%%%%%%%%%%%%% Remark 1
%
%re1 #&#
\begin{rmk}
\label{remarkrmk1}
Assumption \ref{assS}(2) may be restrictive when $q$ is large compared to
$s$, for example, $q=c_nn^{2/5}/\sqrt{\log n}$ with $c_n\to0$ slowly,
$L=c_Ln^{1/5}$ and $s$ bounded. Thus, it would be desirable if
we could replace $r_{qn}$ in the denominators with some quantity
independent of $q$. This is possible in some sense, and here we give
an example. Consider only variable selection, and no
structure identification. Then the penalty
term in the objective function $Q_q(\bgamma)$ is given by $\sum_{j=0}^q
p_\lambda(\| g_j \|_{L_2} )$, and
we assume that $\lambda/\max\{\sqrt{s}r_{sn}, L^{-3/2} \}\to
\infty$. We also
need Assumption \ref{assT}(2) to employ exponential\vspace*{2pt} inequalities and denote
the global minimizer of $l_s(\bgamma_1)=\| y- \gamma_0^T\bB- \sum
_{k=1}^s \gamma_k^T
\bW_k \|_n^2$ on $\mathbb{R}^{(s+1)L}$ by
$\widehat\bgamma_1 \in\mathbb{R}^{(s+1)L}$.\vspace*{2pt} Then, with probability
tending to 1, $(\widehat\bgamma_1^T,
{\mathbf0}^T)^T \in\mathbb{R}^{(q+1)L}$ is a local minimizer of $Q_q(
\overline\bgamma)$, where $\overline\bgamma\in\mathbb
{R}^{(q+1)L}$. %We can prove this fact by following the proof of
%Theorem 1 of \cite{LF2009}.
Thus, some flexibility will
be allowed in the tuning parameter selection when $s$ is bounded.
The proof of this result is outlined in the supplementary material \cite{CHLP2014}. %\ref{suppA}.
%%Section \ref{proofrmk1} of Appendix \ref{proofs}.
\end{rmk}

%%%%%%%%%%%%%%%%%%%%%%%%%%%%%%% proofs

%s4 #&#
\section{Refinement of the group SCAD estimator}\label{secestimation}

%In this section we consider estimation of both the constant and
%time-varying coefficients in semi-varying coefficients model (
%Consider model (\ref{modelsemivarying0}).
To ease the notation, without loss of generality, denote, respectively,
the constant coefficients and the corresponding variables by $\bbeta
_1\in\mathbb{R}^{s_1}$ and ${\mathbf x}_1$, and denote, respectively, the
functional coefficients and the corresponding variables by $\bbeta
_2(t)\in\mathbb{R}^{s_2}$ and ${\mathbf x}_2$. Then we can rewrite model
(\ref{modelsemivarying0}) % for the longitudinal data $(y_i(\bt_i),
as the following:
%
%e4.1 #&#
\begin{eqnarray}\label{modelsemivarying}
y_i(t_{ij}) &=& \beta_0(t_{ij}) + {
\mathbf x}_{1i}(t_{ij})^T \bbeta_1 +{\mathbf
x}_{2i}(t_{ij})^T \bbeta_2(t_{ij})+
\varepsilon_i(t_{ij}),
\nonumber\\[-10pt]\\[-10pt]
\eqntext{j=1,\ldots,m_i,}
\end{eqnarray}
where ${\mathbf x}_{1i}(t_{ij})$ and ${\mathbf x}_{2i}(t_{ij})$ denote,
respectively, the observations on ${\mathbf x}_1(t)$ and ${\mathbf x}_2(t)$ in
the $i$th subject at time $t_{ij}$.
When ${\mathbf x}_1$ and ${\mathbf x}_2$ are given, and $s_1$ and $s_2$ are
fixed and small, this estimation problem has been extensively studied
in the literature \cite{ZFS2009,Li2011}. We revisit this problem to
provide a practical procedure when we encounter ultra-high or large
dimensionality and we do not have a priori knowledge of the relevant
variables, nor which of them have constant coefficients.
First, there is room to improve the group SCAD estimator given in (\ref
{estimatescad}). One reason is that it uses working independence,
which does not hold for longitudinal data in general. %, and it is
%desirable to take into account the correlation structure of the error
%process.
Another reason is that \mbox{B-}spline smoothing suffers from boundary
effects. %Thus we consider (\ref{modelsemivarying}) and refine the
%group SCAD estimator as discussed in this section.
In model (\ref{estimatescad}), the selected variables are divided
into two groups. The variables in~${\mathbf x}_1$ have constant
coefficients with $|(\widehat\beta_k)_c|>0$ and $\|(\widehat\beta
_k)_f\|_{L_2}=0$, and those in ${\mathbf x}_2$ have time-varying
coefficients with $\| (\widehat\beta_k)_f\|_{L_2} > 0$. Note that
when $|(\widehat\beta_k)_c|=0$ and $\| (\widehat\beta_k)_f\|_{L_2}
> 0$ the constant part is zero, but we still include the variable in
${\mathbf x}_2$ without such a constraint on $\beta_k(t)$. % in model (

Our estimation procedure for the coefficients in (\ref
{modelsemivarying}) consists of three steps: (i) constructing initial
estimators, (ii) estimating the covariance function of the error
process and (iii) estimating the coefficients based on the covariance
estimate, which are detailed in the following sections.
Alternatively, after the initial coefficient estimates given in
Section~\ref{sectioninitial} are obtained, we may also iterate
between the covariance function estimation step and the coefficient
estimation step until convergence.

%s4.1 #&#
\subsection{Initial coefficients estimation}\label{sectioninitial}
We could use the group SCAD estimator~(\ref{estimatescad}) as initial
estimator for the coefficients in model (\ref{modelsemivarying}). %
%and take the residuals to estimate the covariance function of the
%error process $\varepsilon(t)$.
However, it may suffer from boundary effects, and the following profile
least squares estimator is \mbox{preferred} \cite{FH2005,LamFan2008}.
Recall that ${\mathbf t}_i=(t_{i1},\ldots,t_{im_i})$, $y_i({\mathbf
t}_i)=(y_i(t_{i1}),\ldots,y_i(t_{im_i}))$, and
$\varepsilon_i ({\mathbf t}_i ) = (\varepsilon_i(t_{i1}), \ldots,\varepsilon
_i(t_{im_i}))$. % are $1\times m_i$ matrices.
Let $K$ denote a kernel function, which is
usually taken as a symmetric p.d.f., and take a bandwidth $h_1>0$.
For any given $\bbeta_1\in\mathbb{R}^{s_1}$, we can estimate $\beta
_0(t)$ and $\bbeta_2 (t)$
in model (\ref{modelsemivarying}) by minimizing the following local
sum of squares:
%
%e4.2 #&#
\begin{eqnarray}\label{estimatebeta2lls}
&& \sum_{i=1}^n%\frac{1}{m_i}
\sum_{j=1}^{m_i} \bigl\{y_i(t_{ij})
-{\mathbf x}_{1i}(t_{ij})^T\bbeta_1-
\bigl(1, {\mathbf x}_{2i}(t_{ij})^T \bigr) \bigl(
\balpha_0 + \balpha_1 (t_{ij} - t) \bigr)
\bigr\}^2K_{h_1}(t_{ij}-t)\hspace*{-17pt}\nonumber
\\
&&\qquad= \sum_{i=1}^n%\frac{1}{m_i}
\bigl
\{y_i({\mathbf t}_i)^T-{\mathbf x}_{1i}(t_{ij})^T
\bbeta_1 - \bigl(\1 _{m_i}, {\mathbf x}_{2i}({\mathbf
t}_i)^T, \bT_i(t) \bigl(\1_{m_i},
{\mathbf x}_{2i}({\mathbf t}_i)^T\bigr) \bigr)\balpha
\bigr\}^T\hspace*{-17pt}
\nonumber\\[-8pt]\\[-8pt]
&&\hspace*{45pt}{}\times \bW_{ih_1}(t) \bigl\{y_i({\mathbf
t}_i)^T-{\mathbf x}_{1i}(t_{ij})^T
\bbeta_1\nonumber
\\
&&\hspace*{96pt}{} - \bigl(\1_{m_i},{\mathbf x}_{2i}({\mathbf
t}_i)^T, \bT_i (t) \bigl(
\1_{m_i}, {\mathbf x}_{2i}({\mathbf t}_i)^T
\bigr) \bigr)\balpha\bigr\},\nonumber
\end{eqnarray}
where $K_{h_1}(\cdot)=K(\cdot/h_1)/h_1$, $\balpha_0,\balpha_1\in
\mathbb{R}^{s_2+1}$, %$\balpha\in\RR^{2(s_2+1)}$,
$\1_{m_i}$ is the $m_i$-dimensional one-vector, $\bT_i=\operatorname{diag}\{
t_{i1}-t,\ldots,t_{im_i}-t\}$,\vspace*{1pt} $\bW_{ih_1}(t)=\operatorname{diag}\{
K_{h_1}(t_{i1}-t),\ldots,K_{h_1}\* (t_{im_i}-t)\}$, and $\balpha
=(\balpha_0^T,\balpha_1^T)^T$.
Let %$\bI_k$ be the $k\times k$ identity matrix and let
$\0_{k\times l}$ be the $k\times l$ dimensional zero matrix.
For any given $\bbeta_1\in\mathbb{R}^{s_1}$, denote the minimizer of
(\ref{estimatebeta2lls}) by $\widetilde\balpha(t,\bbeta_1)$. Then\vspace*{1pt}
an estimator of $(\beta_0(t), \bbeta_2(t)^T)^T$ is %$\big(\widetilde
%
%e4.3 #&#
\begin{eqnarray}\label{estimatebeta2}
&& \bigl(\widetilde\beta_0(t,\bbeta_1),
\widetilde\bbeta_2(t,\bbeta_1)^T
\bigr)^T\nonumber
\\
&&\qquad = (\bI_{s_2+1}, \0_{(s_2+1)\times
(s_2+1)}) \widetilde\balpha(t,\bbeta_1)
\nonumber\\[-8pt]\\[-8pt]
&&\qquad =
(\bI_{s_2+1}, \0_{(s_2+1)\times(s_2+1)}) \bigl(\bV
(t)^T\bW_{h_1}(t)\bV(t)\bigr)^{-1}
\bV(t)^T\bW_{h_1}(t)\nonumber
\\
&&\quad\qquad{}\times (\bY-\bX_{1}
\bbeta_1),\nonumber
\end{eqnarray}
where\vspace*{1pt} $\bY=(y_1({\mathbf t}_1), \ldots, y_1({\mathbf t}_n))^T$, $\bX_1=({\mathbf
x}_{11}({\mathbf t}_1), \ldots, {\mathbf x}_{1n}({\mathbf t}_n))^T$, and $\bV
(t)=(\bV_{1t}^T,\ldots,\bV_{nt}^T)^T$ with $\bV_{it}= (\1
_{m_i},{\mathbf x}_{2i}({\mathbf t}_i)^T, \bT_i(t)(\1_{m_i},{\mathbf x}_{2i}({\mathbf
t}_i)^T) )$, $i=1,\ldots,n$.\vspace*{1pt}

Then, based on working independence, the initial profile least squares
estimator for the constant coefficients $\bbeta_1$ in model (\ref
{modelsemivarying}) is defined as
\[
\widetilde\bbeta_1^{\PLS}=\mathop{\arg\min}_{\bbeta_1\in\mathbb
{R}^{s_1}}
\sum_{i=1}^n%\frac{1}{m_i}
\sum
_{j=1}^{m_i} \bigl\{y_i(t_{ij})
- {\mathbf x}_{1i}(t_{ij})^T \bbeta_1-
\widetilde\beta_0(t_{ij},\bbeta_1) - {\mathbf
x}_{2i}(t_{ij})^T \widetilde
\bbeta_2(t_{ij},\bbeta_1) \bigr
\}^2,
\]
%
%where $(\widetilde\beta_0(t,\bbeta_1),\widetilde\bbeta_2(t,
and the initial estimator for $(\beta_0(t),\bbeta_2(t)^T)^T$ is
defined as $(\widetilde\beta_0^{\PLS}(t),\break\widetilde\bbeta
_2^{\PLS}(t)^T)^T=(\widetilde\beta_0(t,\widetilde\bbeta
_1^{\PLS}), \widetilde\bbeta_2(t,\widetilde\bbeta_1^{\PLS})^T)^T$.
Note that $\widetilde\bbeta_1^{\PLS}$ can be written as
\begin{eqnarray*}
\widetilde\bbeta_1^{\PLS} % \\&=& \argmin_{\bbeta_1\in\RR^{s_1}}
%&=&
&=&
\mathop{\arg\min}_{\bbeta_1\in\mathbb{R}^{s_1}} \bigl\{ ({\mathcal
I}-{\mathcal S}) (\bY-
\bX_1\bbeta_1) \bigr\}^T \bigl\{ ({\mathcal
I}-{\mathcal S}) (\bY-\bX_1\bbeta_1) \bigr\},
\end{eqnarray*}
where\vspace*{1pt} %${\mathcal S}=(\frac{1}{m_1}\bS_{11}^T, \ldots,\frac{1}{m_1}
${\mathcal S}=(\bS_{11}^T, \ldots,\bS_{1m_1}^T,\ldots,\bS
_{n1}^T,\ldots,\bS_{nm_n}^T)^T$, $\bS_{ij}=(1, {\mathbf
x}_{2i}(t_{ij})^T, \0_{1\times(s_2+1)})\* (\bV(t_{ij})^T\bW
_{h_1}(t_{ij})\bV(t_{ij}))^{-1}\bV(t_{ij})^T\bW_{h_1}(t_{ij})$,
$j=1,\ldots,m_i$, $i=1,\ldots,n$, and %${\mathcal I} = \mbox{diag}\{
${\mathcal I} = \bI_{m_1+\cdots+m_n}$.
Thus, we have %by standard least squares theory, $\widetilde
%
%e4.4 #&#
\begin{equation}
\label{estimatebeta1profile} \widetilde\bbeta_1^{\PLS}= \bigl\{
\bX_1^T({\mathcal I}-{\mathcal S})^T({
\mathcal I}-{\mathcal S}) \bX_1 \bigr\}^{-1}
\bX_1^T({\mathcal I}-{\mathcal S})^T({
\mathcal I}-{\mathcal S})\bY,
\end{equation}
and, from the definition of $ (\widetilde\beta
_0^{\PLS}(t),\widetilde\bbeta_2^{\PLS}(t)^T )^T$ and (\ref
{estimatebeta2}), we have
%
%e4.5 #&#
\begin{eqnarray}\label{estimatebeta2profile}
&&\bigl(\widetilde\beta_0^{\PLS}(t),
\widetilde\bbeta_2^{\PLS}(t)^T
\bigr)^T\nonumber
\\
&&\qquad =
(\bI_{s_2+1}, \0_{(s_2+1)\times(s_2+1)}) \bigl(\bV
(t)^T\bW_{h_1}(t)\bV(t)\bigr)^{-1}
\bV(t)^T\bW_{h_1}(t)
\\
&&\quad\qquad{}\times  \bigl(\bY-\bX_1\widetilde
\bbeta_1^{\PLS}\bigr).
\nonumber
\end{eqnarray}
To select the bandwidth $h_1$ in (\ref{estimatebeta1profile}) and
(\ref{estimatebeta2profile}), we choose the value of $h_1$ that
minimizes the leave-one-subject-out cross-validation function.

It is well known that the working independence estimator $\widetilde
\bbeta_1^{\PLS}$ is not semiparametric efficient when the error process
is indeed dependent \cite{wangcarrolllin2005}. In the following
sections, we estimate the covariance function of the error process
using residuals obtained from the initial estimators $\widetilde\bbeta
_1^{\PLS}$, $\widetilde\beta_0^{\PLS}(t)$ and $\widetilde\bbeta
_2^{\PLS}(t)$, and then construct semiparametric efficient estimators.
The semiparametric efficiency results in \cite{Li2011} concern
generalized partially linear models and carry over to the considered
semivarying coefficient models.

%s4.2 #&#
\subsection{Estimation of covariance function of the error process}\label{sectioncovariance}

Denote the covariance function of the error process $\varepsilon(t)$ by
$\phi(u,v)=\cov(\varepsilon(u),\varepsilon(v))$, $u,v\in[0,1]$, and
assume that $\varepsilon(t)$ consists of two independent components:
\[
\varepsilon(t)=\varepsilon_1(t)+\varepsilon_2(t),
\]
where $\varepsilon_1(t)$ has a smooth covariance function $\psi(s,t)$
and $\varepsilon_2(t)$ models the measurement error process. Write the
residuals obtained from the initial profile least squares estimators
$\widetilde\bbeta_1^{\PLS}$ and $\widetilde\bbeta_2^{\PLS}(t_{ij})$
given in Section~\ref{sectioninitial} as
%
%e4.6 #&#
\begin{equation}
\widehat\varepsilon_{ij} = y(t_{ij}) - {\mathbf
x}_{1i}(t_{ij})^T \widetilde
\bbeta^{\PLS}_1- \widetilde\beta_0^{\PLS}(t_{ij})-
{\mathbf x}_{2i}(t_{ij})^T\widetilde
\bbeta_2^{\PLS} (t_{ij}),
\end{equation}
$i=1,\ldots,n, j=1,\ldots,m_i$. We can estimate $\phi$ based on
these residuals.
There exist (semi)parametric approaches to covariance estimation for
longitudinal data \cite{FHL2007,FW2008}. Such methods will be
efficient when the parametric assumptions hold, but can suffer from
large biases otherwise. In general, we may not have knowledge about the
complicated covariance structure and we can use nonparametric methods
to avoid this problem.

Specifically, we use the nonparametric method of \cite{HMW2006} to
estimate the covariance function $\phi$ based on the residuals. % $\hat
First, noting that
\begin{eqnarray}
\phi(t_{ij}, t_{ik})=\psi(t_{ij},
t_{ik})+\var\bigl(\varepsilon_2(t_{ij})
\bigr)I(t_{ij}=t_{ik}),\nonumber
\\
\eqntext{i=1,\ldots,n, j,k=1,\ldots,m_i,}
\end{eqnarray}
we can estimate $\psi(u,v)$ by $\widetilde\psi(u,v) = \widetilde
{a}(u,v)$ where $\widetilde{a}(u,v)$ is the first element of
$\widetilde\ba(u,v)$ which minimizes the following local sum of squares:
%
%e4.7 #&#
\begin{eqnarray}\label{estimatetheta}
&& \sum_{i=1}^n \sum
_{ j\neq k}\large\bigl\{\widehat\varepsilon_{ij}\widehat
\varepsilon_{ik} - a - b (t_{ij}-u) - c(t_{ik}-v)
\large\bigr\} ^2
\nonumber\\[-10pt]\\[-10pt]
&&\hspace*{28pt}{}\times K_{h_2}(t_{ij}-u)K_{h_2}(t_{ik}-v),\nonumber
\end{eqnarray}
%
%&& %\hspace{-30pt} \widetilde{\ba}(u,v)=\argmin_{(a,b,c)^T\in\RR^3}
%&& \hspace{120pt} \times K_{h_2}(t_{ij}-u)K_{h_2}(t_{ik}-v), \nonumber
with a bandwidth $h_2>0$. An explicit formulae for $\widetilde{\psi
}(u,v)$ is available \cite{HMW2006}. % provided an explicit formulae
%for $\widetilde{\psi}(u,v)$:
%(A_1S_{00}-A_2S_{10}-A_3S_{01}),
%where $A_1=S_{20}S_{02}-S_{11}^2$, $A_2=S_{10}S_{02}-S_{01}S_{11}$,
%$A_3=S_{01}S_{20}-S_{10}S_{11}$,
%S_{kl}&=&\frac{1}{N}\sum_{i=1}^n\sum_{j\neq j'} \Big(
%K_{h_2}(t_{ij}-u)K_{h_2}(t_{ij'}-v),\\
%T_{kl} &=& \frac{1}{N}\sum_{i=1}^n\sum_{j\neq j'} \hat\varepsilon(t_{ij})
%and $N=\sum_{i=1}^n m_i(m_i-1)$.
The covariance function estimate $\widetilde\psi(u,v)$ is not
positive semidefinite in general. We can modify this estimate by
truncating the negative components in its spectral decomposition\vspace*{2pt}
$
\widehat\psi(u,v)=\sum_{k=1}^{\zeta_n}\widetilde\omega
_k\widetilde\psi_k(u)\widetilde\psi_k(v)$,
where $\widetilde\omega_1\geq\break \widetilde\omega_2\geq\cdots$ are
the eigenvalues of the operator $\widetilde\psi$, given by
$(\widetilde\psi\alpha)(u)=\break \int_{[0,1]}\alpha(v)\widetilde\psi
(u,v)\,dv$ for $\alpha\in L_2([0,1])$, $\widetilde\psi_j$ is the
eigenfunction corresponding to $\widetilde\omega_j$, $j=1,2,\ldots,$
and $\zeta_n=\max\{k\dvtx \widetilde\omega_j>0, j=1,\ldots,k\}$.

Next, we estimate the variance function of the error process $\varepsilon
(t)$: $\sigma^2(t)\equiv\var(\varepsilon_2(t))$ by $\widehat{\sigma
}{}^2(t)\equiv\widehat{a}$ where $\widehat{a}$ is defined by
%
%e4.8 #&#
\begin{equation}
\label{estimatesigma} \quad (\widehat{a},\widehat{b})^T = \mathop{\arg
\min}_{(a,b)^T\in
\mathbb{R}^2} \sum_{i=1}^n
% \frac{1}{m_i}
\sum_{j=1}^{m_i}\large\bigl\{
\widehat\varepsilon_{ij}^2 - a - b (t_{ij}-t) \large
\bigr\}^2 K_{h_3}(t_{ij}-t),
\end{equation}
with $h_3>0$. Then an estimator for $\phi(u,v)$ is defined as
%
%e4.9 #&#
\begin{equation}
\label{estimatecovariance} \widehat\phi(u,v) = \widehat\psi
(u,v)I(u\neq v) + \widehat{\sigma
}{}^2(u)I(u=v).
\end{equation}
To select the bandwidths $h_2$ and $h_3$ in (\ref{estimatetheta}) and
(\ref{estimatesigma}) we can employ the leave-one-subject-out
cross-validation.

%s4.3 #&#
\subsection{Model estimation accounting for dependent errors}\label{sectioncoefficient}

In this section, we construct semiparametric efficient estimators for
the constant and varying coefficient\vspace*{1pt} functions $\bbeta_1$ and $\bbeta
_2(t)$ using $\widehat\phi(u,v)$ given in (\ref{estimatecovariance}).
Let $\widehat\bLambda_i= (\widehat\phi(t_{ij},t_{ik})
)_{j,k=1,\ldots,m_i}$, $i=1,\ldots,n$. For any $\bbeta_1\in\mathbb
{R}^{s_1}$, define
$\widehat\balpha(t,\bbeta_1)$ as the minimizer of the following
objective function of $\balpha\in\mathbb{R}^{2(s_2+1)}$:
\begin{eqnarray*}
%&& \widehat\balpha(t,\bbeta_1) \\
% &=& \argmin_{\balpha\in\RR^{2(s_2+1)}}
&&\sum_{i=1}^n
\bigl\{y_i({\mathbf t}_i)^T-{\mathbf
x}_{1i}(t_{ij})^T\bbeta_1- \bigl(\1
_{m_i},{\mathbf x}_{2i}({\mathbf t}_i)^T,
\bT_i (t) \bigl(\1_{m_i},{\mathbf x}_{2i}({\mathbf
t}_i)^T\bigr) \bigr)\balpha\bigr\}^T
\widehat{\bLambda}_i^{-1/2}
\\
&&\hspace*{12pt}{}\times \bW_{ih_1}(t)\widehat{\bLambda}_i^{-1/2}
\bigl\{ y_i({\mathbf t}_i)^T-{\mathbf
x}_{1i}(t_{ij})^T\bbeta_1
\\
&&\hspace*{91pt}{} - \bigl(
\1_{m_i},{\mathbf x}_{2i}({\mathbf t}_i)^T,
\bT_i (t) \bigl(\1_{m_i},{\mathbf x}_{2i}({\mathbf
t}_i)^T\bigr) \bigr)\balpha\bigr\}.
\end{eqnarray*}
Then, given $\bbeta_1\in\mathbb{R}^{s_1}$, an estimator for $
(\beta_0(t),\bbeta_2(t)^T )^T$ is taken as %$\big(\widehat
%which can be written as
%
%e4.10 #&#
\begin{eqnarray}\label{estimatebeta2final}
&&\bigl(\widehat\beta_0(t,\bbeta_1),
\widehat\bbeta_2(t,\bbeta_1)^T
\bigr)^T\nonumber
\\
&&\qquad  = (\bI_{s_2+1}, \0_{(s_2+1)\times(s_2+1)}) \widehat
\balpha(t,\bbeta_1)
\nonumber\\[-8pt]\\[-8pt]
&&\qquad =
(\bI_{s_2+1}, \0_{(s_2+1)\times(s_2+1)}) \bigl
(\widehat
\bV(t)^T\bW_{h_1}(t)\widehat\bV(t)\bigr)^{-1}
\widehat\bV(t)^T\bW_{h_1}(t)\widehat\bLambda^{-1/2}\nonumber
\\
&&\quad\qquad{}\times (\bY-\bX_{1}\bbeta_1),
\nonumber
\end{eqnarray}
where $\widehat\bLambda^{-1/2}=\operatorname{diag}\{ \widehat\bLambda
_1^{-1/2}, \ldots, \widehat\bLambda_n^{-1/2}\}$, and $\widehat\bV
(t)=(\widehat\bV_{1t}^T,\ldots,\widehat\bV_{nt}^T)^T$ with
$\widehat\bV_{it}= (\widehat\bLambda_i^{-1/2}(\1_{m_i}, {\mathbf
x}_{2i}({\mathbf t}_i)^T)$, $\widehat\bLambda_i^{-1/2} \bT_i(t)(\1
_{m_i},{\mathbf x}_{2i}({\mathbf t}_i)^T) )$, $i=1,\ldots,n$.%, and recall
%that $\bX_1=(\bx_{11}(\bt_1), \ldots, \bx_{1n}(\bt_n))^T$.

The profile least squares estimator for $\bbeta_1$, denoted by
$\widehat\bbeta_1^{\PLS}$, accounting for within-subject correlation,
is defined as the minimizer of the following objective function of
$\bbeta_1\in\mathbb{R}^{s_1}$:
\begin{eqnarray*}
% && \hspace{80pt} \big\{y_i(\bt_{i})^T -\bx_{1i}(\bt_{i})^T \bbeta_1-{
%&=& \argmin_{\bbeta_1\in\RR^{s_1}}
&&
\sum_{i=1}^n%\frac{1}{m_i}
\sum
_{j=1}^{m_i} \sum_{k=1}^{m_i}
\bigl\{y_i(t_{ij}) -{\mathbf x}_{1i}(t_{ij})^T
\bbeta_1-\widehat\beta_0(t_{ij},
\bbeta_1) -{\mathbf x}_{2i}(t_{ij})^T
\widehat\bbeta_2(t_{ij},\bbeta_1) \bigr
\}^T\widehat\bLambda_i^{-1}(j,k)
\\
&&\hspace*{45pt}{}\times   \bigl\{y_i(t_{ik}) -{\mathbf x}_{1i}(t_{ik})^T
\bbeta_1-\widehat\beta_0(t_{ij},
\bbeta_1) -{\mathbf x}_{2i}(t_{ij})^T
\widehat\bbeta_2(t_{ij},\bbeta_1) \bigr\}.
\end{eqnarray*}
%
%where %${\mathcal X}_2(\bt_i)=\mbox{diag}(\bx_{2i}(t_{i1})^T,\ldots,
%$(\widehat\beta_0(t_{ij},\bbeta_1),\widehat\bbeta_2(t_{ij},
%replaced by $t_{ij}$.
Then the corresponding estimator for $(\beta_0(t),\bbeta_2(t)^T)^T$
is as in (\ref{estimatebeta2final}) with $\bbeta_1$ replaced by
$\widehat\bbeta_1^{\PLS}$. %: $(\widehat\beta_0^{\PLS}(t), \widehat
We call these the refined estimators. Rewrite $\widehat\bbeta
_1^{\PLS}$ as
\begin{eqnarray*}
\widehat\bbeta_1^{\PLS}%\\
% &=& \argmin_{\bbeta_1\in\RR^{s_1}} \sum_{i=1}^n\frac{1}{m_i}
%&& \hspace{70pt} \widehat\bLambda_i^{-1}(j,k)\big\{y_i(t_{ik}) -
%&=& \argmin_{\bbeta_1\in\RR^{s_1}} \sum_{i=1}^n\frac{1}{m_i}
%&& \hspace{120pt}\big\{(1-\widehat\bS_{ik})(y_i(t_{ik}) -
&=&
\mathop{\arg\min}_{\bbeta_1\in\mathbb{R}^{s_1}} \bigl\{ ({\mathcal
I}-\widehat{\mathcal S}) (
\bY-\bX_1\bbeta_1) \bigr\} ^T\widehat
\bLambda^{-1} \bigl\{({\mathcal I}-\widehat{\mathcal S}) (\bY-
\bX_1\bbeta_1) \bigr\},
\end{eqnarray*}
where %${\mathcal I} = \mbox{diag}\{\frac{1}{m_1}\bI_{m_1},\ldots,
$\widehat\bLambda^{-1}=\operatorname{diag}\{\widehat\bLambda_1^{-1},\ldots,\widehat\bLambda_n^{-1}\}$,
%$\widehat{\mathcal S}=(\frac{1}{m_1}\widehat\bS_{11}^T, \ldots,
$\widehat{\mathcal S}\,{=}\,(\widehat\bS_{11}^T, \ldots,\widehat\bS
_{1m_1}^T,\ldots,\widehat\bS_{n1}^T,\ldots,\widehat\bS
_{nm_n}^T)^T$, $\widehat\bS_{ij}=(1,{\mathbf x}_{2i}(t_{ij})^T, \0
_{1\times(s_2+1)}) (\widehat\bV(t_{ij})^T\bW
_{h_1}(t_{ij})\widehat\bV(t_{ij}))^{-1}\widehat\bV(t_{ij})^T\bW
_{h_1}(t_{ij})\widehat\bLambda^{-1/2}$, $j=1,\ldots,m_i$,
$i=1,\ldots,n$. %and ${\mathcal I} = \bI_{m_1+\ldots+m_n}$.
Then we have %by standard least squares theory, $\widehat
%
%e4.11 #&#
\begin{equation}
\label{estimatebeta1refine} \qquad\widehat\bbeta_1^{\PLS}= \bigl\{
\bX_1^T({\mathcal I}-\widehat{\mathcal S})^T
\widehat\bLambda^{-1}({\mathcal I}-\widehat{\mathcal S})
\bX_1 \bigr\}^{-1}\bX_1^T({
\mathcal I}-\widehat{\mathcal S})^T\widehat\bLambda^{-1}({
\mathcal I}-\widehat{\mathcal S})\bY,
\end{equation}
and it follows from the definition of $ (\widehat\beta
_0^{\PLS}(t),\widehat\bbeta_2^{\PLS}(t)^T )^T$ and (\ref
{estimatebeta2final}) that
%
%e4.12 #&#
\begin{eqnarray}\label{estimatebeta2refine}
&& \bigl(\widehat\beta_0^{\PLS}(t),\widehat
\bbeta_2^{\PLS}(t)^T \bigr)^T\nonumber
\\
&&\qquad = (\bI_{s_2+1}, \0_{(s_2+1)\times(s_2+1)})
\\
&&\quad\qquad{}\times \bigl(\widehat\bV(t)^T\bW_{h_1}(t)
\widehat\bV(t)\bigr)^{-1}\widehat\bV(t)^T
\bW_{h_1}(t)\widehat\bLambda^{-1/2}\bigl(\bY-\bX_1
\widehat\bbeta_1^{\PLS}\bigr).
\nonumber
\end{eqnarray}
To select the bandwidth $h_1$ in (\ref{estimatebeta1refine}) and
(\ref{estimatebeta2refine}), we choose the value of $h_1$ that
minimizes the leave-one-subject-out cross-validation function.
% CV(h_1) = \sum_{i=1}^n \frac{1}{m_i} \sum_{j=1}^{m_i} \Big\{
%y_i(t_{ij}) - \bx_{1i}(t_{ij})^T\widehat\bbeta_1^{\PLS,-i} - \widehat
%where $\widehat\bbeta_1^{\PLS,-i}$, $\widehat\beta_0^{\PLS,-i}(t_{ij})$
%and $ \widehat\bbeta_2^{\PLS,-i}(t_{ij})$ are as in (
%$i$-th subject $(y_i(\bt_i), \bx_i(\bt_i))$ removed from the sample.

%%%%%%%%%%%%%%%%%%%%%%%%%%%%%%%%%%%%%%% Simulations

%s5 #&#
\section{Numerical studies}\label{secnumeric}

%s5.1 #&#
\subsection{Simulations}

In our simulation study summarized in this section, the data were
generated from model (\ref{modelsemivarying}),
%$$y_i(\bt)=\beta_0(\bt)+{\mathbf x}_{1i}(\bt)^T{\bolds\beta}_1+{\mathbf x}_{2i}(
where each ${\mathbf t}_i$ is a vector of $m$ equally-spaced grid points on
$[0,1]$. We considered three coefficients settings:
\begin{longlist}[\textit{Case} III.]
\item[\textit{Case} I.] $\beta_0(t)=3.5\sin(2\pi t)$, $s_1=2$, $s_2=3$, ${\bolds
\beta}_1=(5,-5)^T$ and  ${\bolds\beta}_2(t)=(5(1-t)^2, 3.5(\exp
(-(3t-1)^2)+\exp(-(4t-3)^2))-1.5, 3.5t^{1/2})^T$.

\item[\textit{Case} II.] $\beta_0(t)=3.5\sin(2\pi t)$, $s_1=0$, $s_2=5$, and
${\bolds\beta}_2(t)=(5(1-t)^2$,\break   $3.5(\exp(-(3t-1)^2)+\exp
(-(4t-3)^2))-1.5, 3.5t^{1/2}, 6-2t, 2-3\cos(4\pi t))^T$.

\item[\textit{Case} III.] $\beta_0(t)=3.5\sin(2\pi t)$, $s_1=5$, $s_2=0$,
${\bolds\beta}_1=(5,-5,2.5,-2.5,1)^T$.
\end{longlist}
%
%Case II is the varying-coefficient model and Case III is the partially
%linear model. These two cases are included in the simulation study to
%demonstrate the capability of the group SCAD estimator in identifying
%the true model structure even without knowing whether $s_1=0$ and
%whether $s_2=0$.

We generated covariates ${\mathbf x}(t)$ from a $p$-dimensional Gaussian
process whose component processes each has mean zero and covariance
function $\cov(x_{k}(s),\break  x_{k}(t))= 2\sin(2\pi s)\sin(2\pi t)$.
The correlation between components is specified as follows. The first
$s_1+s_2+s_0$ elements of ${\mathbf x}(t)$ are correlated with a constant
correlation $\rho$, and thus follow a compound-symmetry covariance
structure. The remaining $p-(s_1+s_2+s_0)$ elements of ${\mathbf x}(t)$ are
uncorrelated with each other and the first $s_1+s_2+s_0$ elements. The
first $s_1$ and $s_2$ elements of ${\mathbf x}(t)$ are used as ${\mathbf
x}_{1}(t)$ and ${\mathbf x}_{2}(t)$ in the model, respectively. The next
$s_0$ elements of ${\mathbf x}(t)$ are spurious variables that are not
related to $y(t)$ but correlated to ${\mathbf x}_{1}(t)$ and ${\mathbf
x}_{2}(t)$. The random\vadjust{\goodbreak} error $\varepsilon(t)$ was simulated from an
ARMA($1,1$) Gaussian process with mean zero and covariance function
$\cov(\varepsilon(s),\varepsilon(t))= {\omega} {\mathrm r}^{|s-t|}$,
with ${\omega}=0.85$ and ${\mathrm r}=0.5$. %We set ${\omega}=0.85$ and ${

In addition, we considered two more cases with different covariate and
error distributions.
\begin{longlist}[\textit{Case} IV.]
\item[\textit{Case} IV.] The\vspace*{1pt} same as case~I except that the covariance matrix
of the first $s_1+s_2+s_0$ elements of ${\mathbf x}(t)$ is an AR(1) matrix
with $(j,j')$th element equal to $\rho^{|j-j'|}$. We set ${\omega
}=0.85$ and ${\mathrm r}=0.6$ for the error process.

\item[\textit{Case} V.] The same as case~I except that the covariance matrix of
the first $s_1+s_2+s_0$ elements of ${\mathbf x}(t)$ is a symmetric matrix
with $(j,j')$th element equal to $|j-j'|/\{2(s_1+s_2+s_0)\}+\rho
^{|j-j'|}$. We set ${\omega}=0.95$ and ${\mathrm r}=0.5$ for the error process.
\end{longlist}

%
%t1 #&#
\begin{sidewaystable}%[htp]
\tabcolsep=0pt
\tablewidth=\textwidth
\caption{Variable selection results from 500 simulations. $\Cvar$ (Cfix)
is the proportion of variables with varying-coefficients (constant
nonzero coefficients) that are selected; %Cfix is the proportion that
%variables with constant nonzero coefficients are all selected;
Size is the average model size; U (O) is the proportion of
underselection (overselection); % O is the proportion of overselection;
TP (FP) is the average number of true positive (false positive); %FP is
%the average number of false positive;
TPvar (FPvar) is the average number of true positive (false positive)
for the varying-coefficients; % (identified variables with
%varying-coefficients indeed have varying-coefficients); %FPvar is the
%average number of false positive for varying-coefficient;
TPfix (FPfix) is the average number of true positive (false positive)
for the constant coefficients; % (identified variables with constant
%nonzero coefficients indeed have constant nonzero coefficients);
%FPvar is the average number of false positive for constant coefficient.
MMMS is the median of the minimum model size to contain all true
nonzeros in the screening step.
Here, $m=20,s_0=10,p=500$, $n$ is the sample size and $\rho$ is the
spurious correlation. The values in the parentheses are the robust
standard deviations}\label{table-1}
\begin{tabular*}{\tablewidth}{@{\extracolsep{\fill}}@{}ld{1.8}d{1.8}d{1.8}d{1.8}d{1.8}d{1.8}d{1.8}d{1.8}d{1.8}d{1.8}d{1.8}@{}}
\hline
\textbf{Case} &\multicolumn{3}{c}{\textbf{I}}& \multicolumn{2}{c}{\textbf{II}}& \multicolumn{2}{c}{\textbf{III}} & \multicolumn{2}{c}{\textbf{IV}}& \multicolumn{2}{c@{}}{\textbf{V}}\\[-6pt]
\hrulefill &\multicolumn{3}{c}{\hrulefill}& \multicolumn{2}{c}{\hrulefill}& \multicolumn{2}{c}{\hrulefill} & \multicolumn{2}{c}{\hrulefill}& \multicolumn{2}{c@{}}{\hrulefill}
\\
$\bolds{n}$ & \multicolumn{1}{c}{\textbf{100}} & \multicolumn{1}{c}{\textbf{100}}
& \multicolumn{1}{c}{\textbf{200}} & \multicolumn{1}{c}{\textbf{100}}
& \multicolumn{1}{c}{\textbf{200}}& \multicolumn{1}{c}{\textbf{100}}
& \multicolumn{1}{c}{\textbf{200}}&\multicolumn{1}{c}{\textbf{100}}
& \multicolumn{1}{c}{\textbf{200}} & \multicolumn{1}{c}{\textbf{100}} & \multicolumn{1}{c@{}}{\textbf{200}}
\\
$\bolds{\rho}$ & \multicolumn{1}{c}{\textbf{0.1}} & \multicolumn{1}{c}{\textbf{0.5}}
& \multicolumn{1}{c}{\textbf{0.5}} & \multicolumn{1}{c}{\textbf{0.3}} & \multicolumn{1}{c}{\textbf{0.3}}
& \multicolumn{1}{c}{\textbf{0.4}} & \multicolumn{1}{c}{\textbf{0.4}} & \multicolumn{1}{c}{\textbf{0.4}}
& \multicolumn{1}{c}{\textbf{0.5}} & \multicolumn{1}{c}{\textbf{0.4}} & \multicolumn{1}{c@{}}{\textbf{0.5}}\\
\hline
$\Cvar$ & 0.965 & 0.904 & 0.996&0.952 & 0.986 &\multicolumn{1}{c}{--} & \multicolumn{1}{c}{--}&0.976 &0.992 &0.925 &0.998\\
Cfix & 0.926& 0.812 & 0.912&\multicolumn{1}{c}{--} &\multicolumn{1}{c}{--}& 0.938& 0.969 &0.854 &0.936 &0.810&0.914\\
Size & 5.01~(0.79) &6.43~(1.23) & 5.02~(0.54)&5.12~(0.13) & 5.04~(0.01)& 5.25~(0.13)& 5.01~(0.01) & 4.95~(0.63) & 4.98~(0.40) & 4.87~(0.96) & 4.99~(0.22) \\
U & 0.00 & 0.05 & 0.00&0.00& 0.00& 0.00& 0.00 &0.02 &0.01 &0.03 &0.00\\
O &0.01& 0.19 & 0.02&0.08& 0.01& 0.11& 0.03 &0.03 &0.00 &0.04 &0.00\\
TP & 5.00~(0.79) & 4.99~(1.18) &5.00~(0.54) & 4.97~(0.13) & 5.00~(0.01) &4.96~(0.12)& 5.00~(0.01) & 4.93~(0.59) & 4.97~(0.40) & 4.81~(0.93) & 4.99~(0.22) \\
FP & 0.01~(0.06) & 0.81~(0.25) & 0.02~(0.01)& 0.09~(0.01)& 0.04~(0.00)& 0.15\mbox{~}(0.01) & 0.05~(0.01) &0.03~(0.19) &0.00~(0.00) &0.05~(0.19) &0.00~(0.00)\\
TPvar &2.93~(0.54) & 2.87~(0.75) & 2.97~(0.42)& 4.98~(0.11)& 5.00~(0.04)& \multicolumn{1}{c}{--}&\multicolumn{1}{c}{--} & 2.82~(0.48) & 2.92~(0.34) & 2.79~(0.65) & 2.97~(0.31)\\
FPvar &0.10~(0.04) & 0.15~(0.09) & 0.09~(0.06)& 0.03~(0.08) & 0.01~(0.00)
&0.03~(0.08) &0.01~(0.01) &0.01~(0.08) &0.00~(0.00) &0.04~(0.21) &0.01~(0.05)\\
TPfix&1.92~(0.33) & 1.84~(0.48) & 1.92~(0.22)& \multicolumn{1}{c}{--}& -& 4.93~(0.15) & 5.00~(0.04)& 1.96~(0.26) & 1.98~(0.17) & 1.88~(0.43) & 1.99~(0.09)\\
FPfix&0.04~(0.28) & 0.80~(0.42) & 0.14~(0.21) &0.06~(0.14) &0.03~(0.04)& 0.10~(0.12) & 0.02~(0.01) &0.16~(0.40) &0.06~(0.23) &0.20~(0.42) &0.08~(0.27)
\\[6pt]
MMMS & \multicolumn{1}{l}{5~(0)} & \multicolumn{1}{l}{5~(0)} & \multicolumn{1}{l}{5~(0)} &
\multicolumn{1}{l}{5~(0)} & \multicolumn{1}{l}{5~(0)} & \multicolumn{1}{l}{5~(0)} &
\multicolumn{1}{l}{5~(0)} & \multicolumn{1}{l}{5~(0)} & \multicolumn{1}{l}{5~(0)} &
\multicolumn{1}{l}{5~(0)} & \multicolumn{1}{l@{}}{5~(0)}\\
\hline
\end{tabular*}
\end{sidewaystable}

%
%t2 #&#
\begin{table}%[htbp]
\tabcolsep=0pt
\caption{Estimation results of 500 simulations. Oracle (Practical)
estimate refers to estimation with known model (after screening and
model selection)} \label{table-2}
\begin{tabular*}{\tablewidth}{@{\extracolsep{\fill}}@{}lcccccccc@{}}
\hline
&\multicolumn{4}{c}{\textbf{Oracle estimate}}&\multicolumn{4}{c@{}}{\textbf{Practical estimate}}\\[-6pt]
&\multicolumn{4}{c}{\hrulefill}&\multicolumn{4}{c@{}}{\hrulefill}
\\
&\multicolumn{2}{c}{\textbf{Initial} $\bolds{\widetilde\beta_k^{\PLS}}$}&\multicolumn{2}{c}{\textbf{Refined} $\bolds{\widehat\beta_k^{\PLS}}$}
&\multicolumn{2}{c}{\textbf{Initial} $\bolds{\widetilde\beta_k^{\PLS}}$}&\multicolumn{2}{c@{}}{\textbf{Refined} $\bolds{\widehat\beta_k^{\PLS}}$}
\\[-6pt]
&\multicolumn{2}{c}{\hrulefill}&\multicolumn{2}{c}{\hrulefill} &\multicolumn{2}{c}{\hrulefill} & \multicolumn{2}{c@{}}{\hrulefill}
\\
\multicolumn{9}{@{}l@{}}{\textbf{Case I}}\\
\textbf{Parameters} & \textbf{MAE} & \textbf{RMSE} & \textbf{MAE} & \textbf{RMSE} & \textbf{MAE} & \textbf{RMSE} & \textbf{MAE} & \textbf{RMSE}\\
\hline
$\beta_{11}$ &0.0374& 0.0623& 0.0253& 0.0387 & 0.0572& 0.0806& 0.0266&0.0458\\
$\beta_{12}$ &0.0507& 0.0642& 0.0296& 0.0417 & 0.0662 & 0.0768&0.0330& 0.0500
\\[6pt]
\textbf{Functions} & \textbf{MIAE} & \textbf{RMISE} & \textbf{MIAE} & \textbf{RMISE} & \textbf{MIAE} & \textbf{RMISE} & \textbf{MIAE} & \textbf{RMISE}\\
\hline
$\beta_0(t)$ &0.1678& 0.2410& 0.0872 & 0.1526 & 0.1922& 0.2863&0.1020& 0.1902\\
$\beta_{21}(t)$ &0.1697& 0.2497 & 0.1098& 0.1805 & 0.2066& 0.2819&0.1243& 0.2111\\
$\beta_{22}(t)$ &0.1526& 0.2433& 0.1151 & 0.1568 & 0.1815& 0.2760&0.1261& 0.1957\\
$\beta_{23}(t)$ &0.1804& 0.2819 & 0.1241& 0.2007 & 0.2119& 0.2939&0.1317& 0.2293
\\[6pt]
\multicolumn{9}{@{}l@{}}{\textbf{Case II}}\\
\textbf{Functions} & \textbf{MIAE} & \textbf{RMISE} & \textbf{MIAE} & \textbf{RMISE} & \textbf{MIAE} & \textbf{RMISE} & \textbf{MIAE} & \textbf{RMISE}\\
\hline
$\beta_0(t)$ &0.1748& 0.2571& 0.1042 & 0.1794 & 0.2254& 0.3179&0.1220& 0.2535\\
$\beta_{21}(t)$ &0.1939& 0.2703 & 0.0859& 0.1389 & 0.2316& 0.3315&0.1015& 0.2220\\
$\beta_{22}(t)$ &0.1532& 0.2357& 0.1029 & 0.1473 & 0.1964& 0.3003&0.1221& 0.2088\\
$\beta_{23}(t)$ &0.1710& 0.2381 & 0.1019& 0.1462 & 0.2156& 0.2901&0.1215& 0.2383\\
$\beta_{24}(t)$ &0.2074 & 0.3352& 0.1181 & 0.1889& 0.2473& 0.3880&0.1243& 0.2565\\
$\beta_{25}(t)$ &0.2425 & 0.3562& 0.1252 & 0.2441& 0.2680& 0.4055&0.1362& 0.2590
\\[6pt]
\multicolumn{9}{@{}l@{}}{\textbf{Case III}}\\
\textbf{Parameters} & \textbf{MAE} & \textbf{RMSE} & \textbf{MAE} & \textbf{RMSE} & \textbf{MAE} & \textbf{RMSE} & \textbf{MAE} & \textbf{RMSE} \\
\hline
$\beta_{11}$ &0.0136& 0.0264& 0.0109& 0.0223 & 0.0142& 0.0402& 0.0136&0.0387\\
$\beta_{12}$ &0.0125& 0.0200& 0.0118& 0.0141 & 0.0144 & 0.0458&0.0138& 0.0374\\
$\beta_{13}$ &0.0256&0.0400& 0.0162&0.0332&0.0352& 0.0538& 0.0286&0.0469\\
$\beta_{14}$ &0.0206&0.0360& 0.0175&0.0282&0.0327& 0.0565&0.0249&0.0489\\
$\beta_{15}$ & 0.0300&0.0400 & 0.0265&0.0346&0.0525& 0.0648& 0.0430&0.0608
\\[6pt]
\textbf{Functions} & \textbf{MIAE} & \textbf{RMISE} & \textbf{MIAE} & \textbf{RMISE} & \textbf{MIAE} & \textbf{RMISE} & \textbf{MIAE} & \textbf{RMISE}\\
\hline
$\beta_0(t)$ &0.1067& 0.0985& 0.1038 & 0.0938 & 0.1215& 0.1126&0.1156& 0.1101\\
\hline
\end{tabular*}
\end{table}

After 500 simulations, we summarized the performance of our variable
selection and structure identification procedures in Table~\ref
{table-1} for the five considered cases. The results for larger values
of $\rho$ under cases II and III are very similar to that under case
I, and thus are not reported here. Both varying-coefficients and
constant nonzero coefficients were selected correctly with high
probability over the simulations, especially when the spurious
correlation $\rho$ is low. True positive fractions for the overall
model, the varying-coefficient part and the constant coefficient part
were close to the true number of variables while false positive
fractions were small in general. When spurious correlation was moderate
or high, under-selection and over-selection were observed more often.
In general, the selection accuracy was improved as we increased the
sample size $n$.

\begin{figure}[t!]

\includegraphics{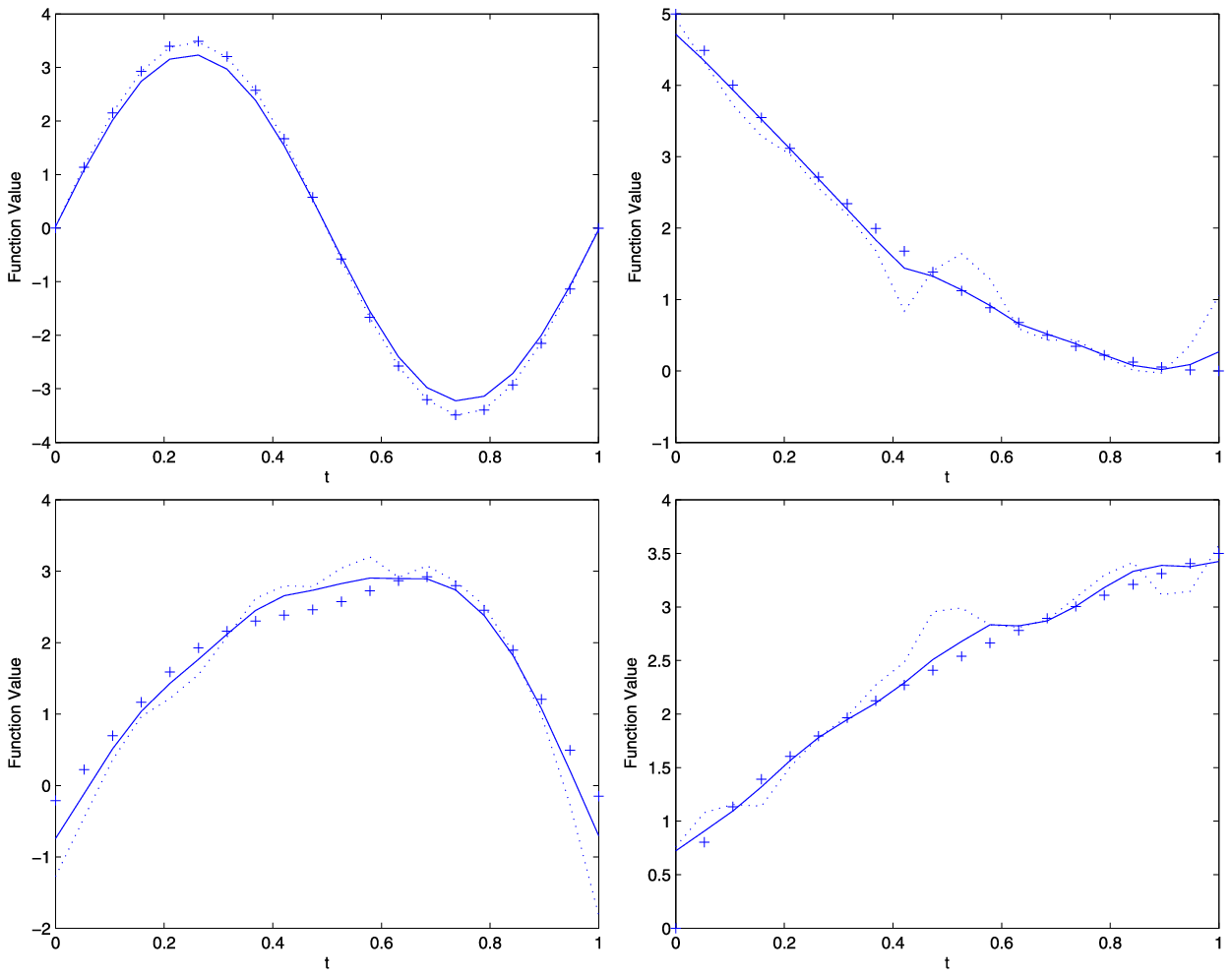}

\caption{Estimated varying coefficients with median MISE for case~I.
%%The top two panels are for
From left to right are (top) $\beta_0(t)$, $\beta_{21}(t)$, (bottom) $\beta
_{22}(t)$ and $ \beta_{23}(t)$. The lines with ``$+$'' symbols are the
true functions; the solid and dashed lines are, respectively, the refined
and initial estimates.}\label{fig0}\vspace*{12pt}
\end{figure}

The estimation results for various model components are summarized in
Table~\ref{table-2} for the three cases I--III with $n=100$, $m=20$ and
$\rho=0.1$. The results for
moderate or higher correlation values are similar, and are thus skipped
here. For estimates of the parametric components, we report the
estimation mean absolute error (MAE) and the root mean square error
(RMSE). For estimates of the nonparametric components, we report the
mean integrated absolute error (MIAE) and the root mean integrated
squared error (RMISE). The practical estimates performed almost as well
as the respective oracle estimates. The refined estimates in general
performed better than the respective initial estimates. Typical
estimates for the coefficient functions in case~I with median MISE are
depicted in Figure~\ref{fig0}.

%
%f1 #&#

%
%f2 #&#
\begin{figure}[t!]

\includegraphics{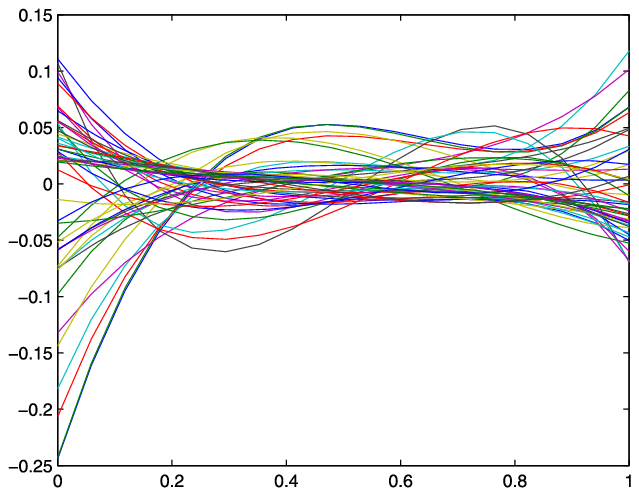}

\caption{Estimated varying coefficients of the 51 TFs obtained from
NIS.}\label{figsc}
\end{figure}

%s5.2 #&#
\subsection{Real data analysis}\label{yeast}
We analyzed the well-known Yeast Cell Cycle gene expression data set,
originally studied by \cite{Spellman1998}. There are $n=297$
cell-cycle-regularized genes whose expression levels were measured at
$m=18$ time points covering two cell-cycle periods. We aim at
identifying important transcription factors (TF) that affect the gene
expression. Using the same subset of the original data as in \cite
{WCL2007}, we included totally $p=96$ TFs as covariates in the
following analysis.

We first applied the nonparametric screening procedure and 51 TFs were
kept after the screening. The nonparametric estimates of
varying-coefficients for the 51~TFs from the inital screening are
plotted in Figure~\ref{figsc}. The names of these TFs are given in the
following list:\vspace*{9pt}

{\small
\begin{verbatim}
HIR2 HIR1 MET4 FKH2 NDD1 SWI4 SWI5 SKN7 FKH1 MCM1
SMP1 PHD1 SWI6 PUT3 ACE2 MBP1 CIN5 ABF1 RLM1 GRF10.Pho2.
MSN1 RTG1 STE12 SOK2 RGM1 MTH1 CBF1 RTG3 STB1 INO4
DOT6 GAT3 SIP4 REB1 STP1 YAP6 HAL9 DAL81 GAL4 YAP5
PDR HAP4 MSN4 RAP1 DIG1 CUP9 NRG1 INO2 HAP5 FHL1 RFX1
\end{verbatim}
}\vspace*{9pt}
The above list includes all of the TFs mentioned in \cite{WLH2008}.
Comparing to the 21 known yeast TFs related to the cell cycle process
included in Figure~2 of \cite{WCL2007}, our list does not include
BAS1, GCN4, GCR1, GCR2, LEU3 and MET31 and includes all of the other 15
TFs. Comparing to the additional TFs identified in Table~2 of~\cite{WCL2007}, the above list does not include 23 of their totally 52 TFs.

%
%f3 #&#
\begin{figure}%[b!]
\begin{tabular}{@{}cccc@{}}

\includegraphics{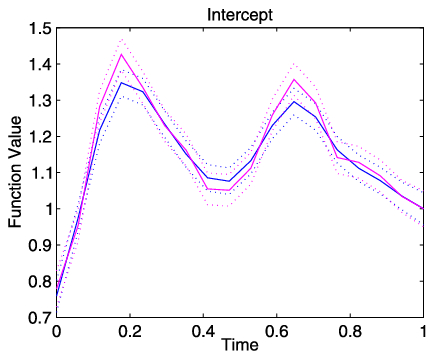}
 & \includegraphics{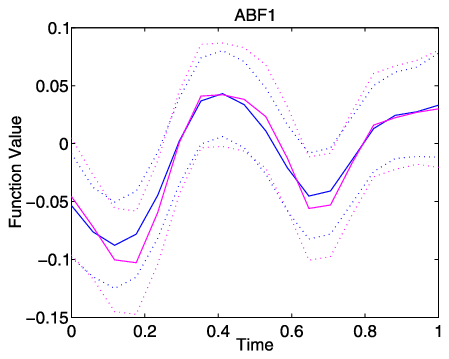}\\[6pt]

\includegraphics{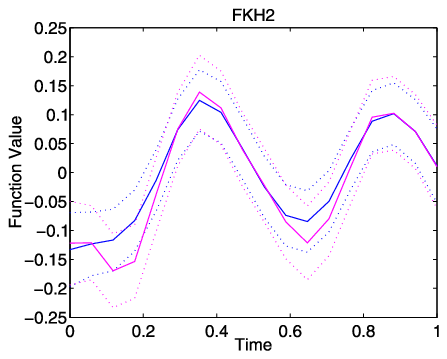}
 & \includegraphics{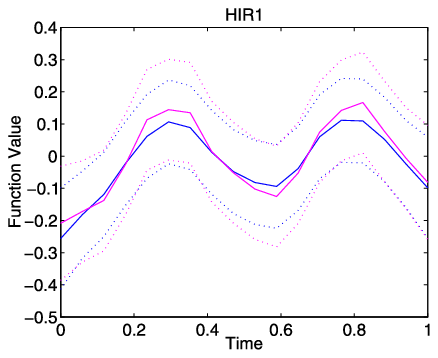}\\[6pt]

\includegraphics{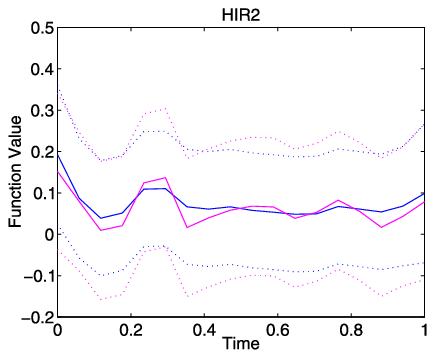}
 & \includegraphics{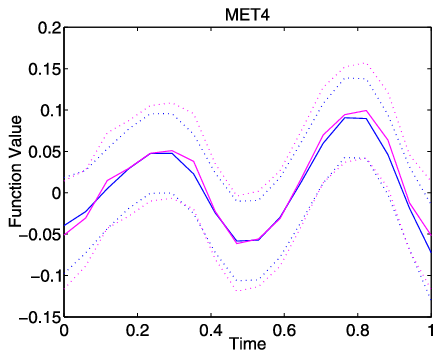}%\\[6pt]
\end{tabular}
\caption{Estimated varying coefficients for Yeast Cycle data. Blue
curves display the refined estimates and red curves display the initial
estimates. Solid curves are the estimated functions while dashed curves
are the 95\% confidence intervals.}
\label{fig1}
\end{figure}

\setcounter{figure}{2}
%
%f3 #&#
\begin{figure}%[b!]
\begin{tabular}{@{}cccc@{}}

\includegraphics{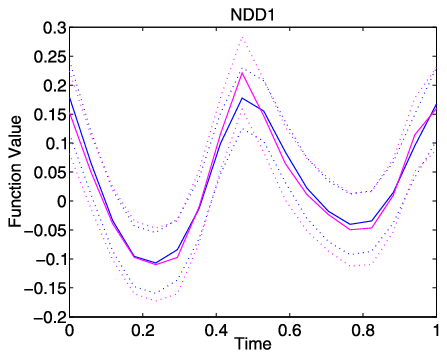}
 & \includegraphics{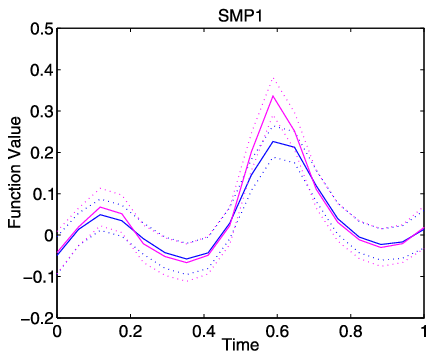}\\[6pt]

\includegraphics{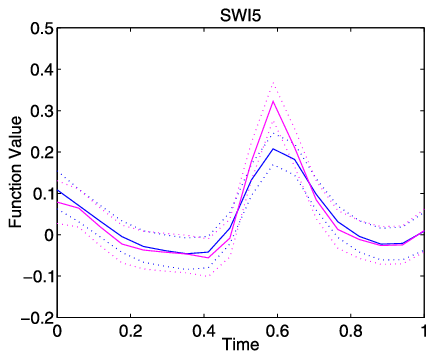}
 & \includegraphics{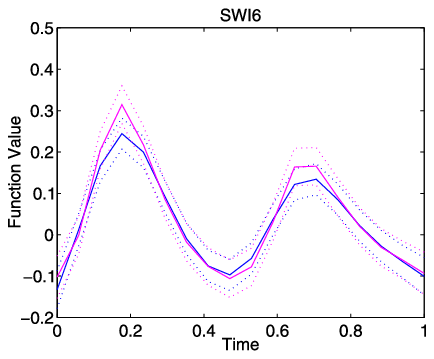}
\end{tabular}
\caption{(Continued).}
\end{figure}

%
%t3 #&#
\begin{table}%[htbp]
\tabcolsep=0pt
\tablewidth=210pt
\caption{Estimated constant coefficients for two transcription factors
in the Yeast Cell Cycle data. In the parentheses are the bootstrap
standard errors} \label{table-3}
\begin{tabular*}{\tablewidth}{@{\extracolsep{\fill}}@{}lcc@{}}
\hline
\textbf{TF} & \textbf{Initial} $\bolds{\widetilde\beta_k^{\PLS}}$ & \textbf{Refined} $\bolds{\widehat\beta_k^{\PLS}}$\\
\hline
MCM1& \phantom{$-$}0.0129 (0.0103) & \phantom{$-$}0.0220 (0.0101)\\
RLM1& $-$0.0032 (0.0095) & $-$0.0097 (0.0094)\\
\hline
\end{tabular*}
\end{table}

We then applied our proposed variable selection and structure
identification procedure and identified 11 TFs, among which 9 TFs have
varying-coefficients and the other 2 TFs have constant coefficients.
The refined estimates for the 9 varying-coefficients are plotted in
Figure~\ref{fig1}, along with the 95\% confidence intervals computed
from bootstrap based on 500 resamples. For the sake of comparison, in
each panel we also display the corresponding initial estimate and its
95\% bootstrap confidence interval. The two estimates are similar but
have distinctions. The confidence intervals for the initial estimates
are always slightly wider than those for the respective refined
estimates. The estimated constant coefficients are given in Table~\ref
{table-3} where the standard errors (SE) were computed based on the bootstrap.

Our results are comparable to previous publications. The estimated
varying-coefficients almost all show periodic transcriptional effects,
as was evidenced in earlier publications. Of the 9 TFs with varying
coefficients, SWI6, FKH2, NDD1 and SWI5 were also identified as
important TFs in \cite{WCL2007} and \cite{WLH2008}; ABF1, HIR1, HIR2,
MET4 and SMP1 were also identified as important TFs in \cite{WCL2007}.
Of the 2~TFs with constant coefficients, MCM1 was identified before in
\cite{WCL2007} and \cite{WLH2008} but its effect was estimated as a
varying coefficient instead of a constant one; RLM1 was also included
in the list of important TFs reported in \cite{WCL2007}. Furthermore,
\cite{WZQ2012}~used a penalized estimating equation approach to
analyze this data set and identified similar number of TFs although the
authors did not report the names of the identified~TFs.

\begin{figure}%[h]%[b!]

\includegraphics{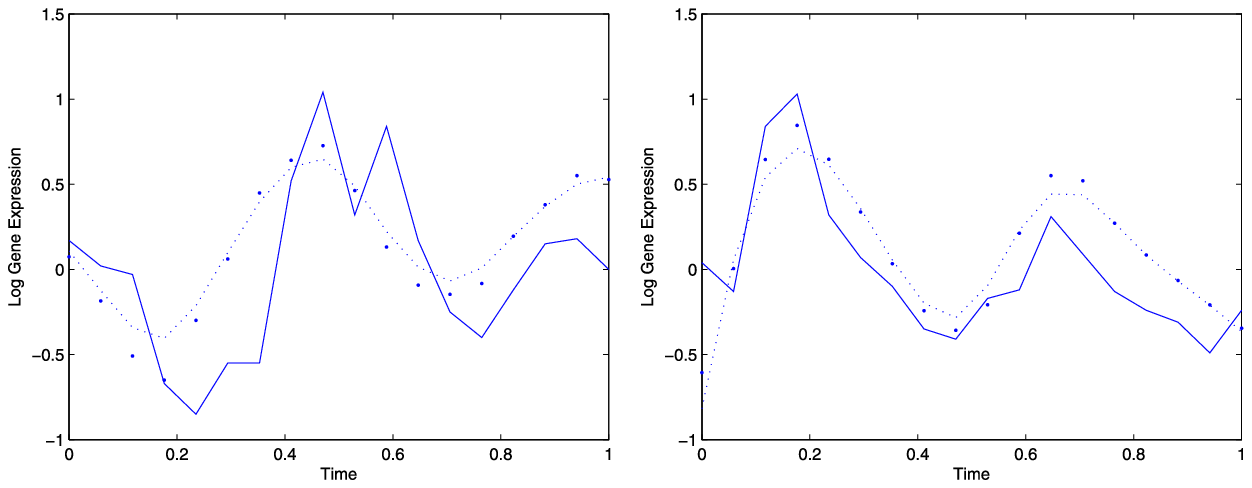}

\caption{Observed and fitted response curves for subjects 25 (left)
and 177 (right) in the sample. Solid lines are the observed response
curves; broken lines are the fitted curves from refined estimates;
Dotted lines are the fitted curves from the initial estimates.}\label{fig6}
\end{figure}

For two typical individuals selected from the data set, we displayed
their observed and fitted time-varying responses in Figure~\ref{fig6}.
The prediction from the fitted model resembles the true functional
response closely and provides a more natural and smooth interpretation
for the cell cycle process. These results may serve as useful tools for
biologists to study molecular events with large variability.

We remark that the cell cycle is a complicated biological process and
the between-gene heterogeneity may prohibit investigators from making
the identical distribution assumption. Our endeavor here is to model a
collective time-varying effect of the TF that remains relatively fixed
among genes. A~more refined analysis for individual phases of the cycle
may be carried out to reduce the variability. Caution must be exercised
to generalize the results, especially to a set of genes with entirely
different regulatory mechanisms.

%
%f4 #&#

%%%%%%%%%%%%%%%% APPENDIX
\begin{appendix}\label{secproofgs}
%s6 #&#
\section*{Appendix: Proofs of Theorems \texorpdfstring{\lowercase{\protect\ref{teothm3}--\protect\ref{teothm5}}}{3-1--3.3}} %

%%%%%%%%%%%%%%%%%%%%%%%%%%%%%%%%%%%%%%%%%%%%%% gSCAD
%%%%%%%%%%%%%%%%%%%%%%%%%%%%%%%%%%%%%%%%%%%%%% Proofs of Theorems

%In this section we prove Theorems \ref{teothm3}-\ref{teothm5}.
We describe some important facts first. Recall that $|\bgamma|$
denotes the Euclidean norm of $\bgamma$.
It is easy to see that, for some positive
$C_{N1}$ and $C_{N2}$,
%
%e6.1 #&#
\begin{equation}
C_{N1} L^{-1/2}|\bgamma| \le\bigl\|\bigl(\gamma_0^T
\bB, \ldots, \gamma_q^T\bB\bigr)^T
\bigr\|_{L_2} \le C_{N2} L^{-1/2}|\bgamma| \label{eqnh598}
\end{equation}
uniformly in $\bgamma=(\gamma_0^T, \ldots, \gamma_q^T)^T$. Lemmas 3
and 4 %\ref{lemlem3} and \ref{lemlem4}
of the supplementary material~\cite{CHLP2014} imply that
there are positive constants $C_{B1}$ and $C_{B2}$ such that, with
probability tending to 1,
\[
C_{B1}/L \le\lambda_{\min}\bigl( \bigl\langle\bB,
\bB^T \bigr\rangle_n \bigr) \le\lambda_{\max}
\bigl( \bigl\langle\bB, \bB^T \bigr\rangle_n \bigr) \le
C_{B2}/L.
\]

%s6.1 #&#
\subsection{Proof of Theorem \texorpdfstring{\protect\ref{teothm3}}{3.1}}
%%%%%%%%%%%%%%%%%%%%%%%%%%%%% Proof of Theorem 3

%%%%%%%%%%%%%%%%%%%%%%%%%%%%%%%%%%%%%%%%%%%% Def. of \blambda_m
Define $\bgamma_m=(\gamma_{0m}^T, \ldots,
\gamma_{qm}^T)^T \in R^{(q+1)L}$ by
%
%e6.2 #&#
\begin{equation}
\bgamma_m = \mathop{\arg\min}_{(\gamma_{0}^T, \ldots,
\gamma_{q}^T)^T \in R^{(q+1)L}} \Biggl\|
\beta_0+\sum_{k=1}^q
\beta_kx^{(k)} -\gamma_0^T\bB- \sum
_{k=1}^q\gamma_k^T
\mathbf{W}_{k} \Biggr\|_n^2. \label{eqnh601}
\end{equation}
Lemma 5 %\ref{lemlem5}
of the supplementary material \cite{CHLP2014} implies that $l_q( \bgamma)$
is strictly concave with probability tending to 1. Thus, by\vspace*{2pt}
Lemma 6 %\ref{lemlem6}
of the supplementary material \cite{CHLP2014}, we have only to demonstrate that
there is a local minimizer of $l_q( \bgamma)$ on $R^{(q+1)L}$, denoted by
$\widetilde\bgamma= (\widetilde\gamma_0^T,\ldots, \widetilde\gamma_q^T)^T$, satisfying
\[
\bigl\| \bigl( \widetilde\gamma_0^T\bB, \ldots, \widetilde
\gamma_q^T\bB\bigr)^T - \bigl(
\gamma_{0m}^T\bB, \ldots, \gamma_{qm}^T
\bB\bigr)^T \bigr\|_{L_2}^2 = O_p
\bigl(r_{qn}^2\bigr).
\]
Recalling the definition of $r_{qn}$, given in (\ref{eqnh503}), and
(\ref{eqnh598}), we
define $\Gamma_M$ by
\[
\Gamma_M = \bigl\{ \bgamma\in R^{(q+1)L} | |\bgamma-
\bgamma_m| = M(qL/n)^{1/2}L^{1/2} \bigr\},
\]
for a positive $M$, and represent $l_q( \bgamma)$ as
%
%e6.3 #&#
\begin{eqnarray}\label{eqnh611}
l_q (\bgamma) &=& l_q(\bgamma_m
) -2(\bgamma-\bgamma_m)^T \bigl\langle\bigl(
\bB^T \town^T\bigr)^T, \varepsilon\bigr
\rangle_n
\nonumber\\[-8pt]\\[-8pt]
&&{} + (\bgamma-\bgamma_m)^T\Sigma_n (\bgamma-\bgamma_m).\nonumber
\end{eqnarray}
%
%l_q (\bgamma) &= l_q(\bgamma_m ) -2(\bgamma-\bgamma_m)^T
%& \qquad
%+ (\bgamma-\bgamma_m)^T
% \nip{(\bB^T \town^T)^T, (\bB^T \town^T)}
%& = l_q( \bgamma_m ) + J_1(\bgamma) + J_2 (\bgamma)\quad
%(say).\notag
By Lemma 7 % \ref{lemlem7}
of the supplementary material \cite{CHLP2014}, we have uniformly on $\Gamma_M$, the first term
in the right-hand side of (\ref{eqnh611}) is $MqLn^{-1}O_p(1)$.
%J_1 (\bgamma) = MqLn^{-1}O_p(1).
By Lemma 5 % \ref{lemlem5}
of the supplementary material \cite{CHLP2014}, we have the second term in the right-hand side of
(\ref{eqnh611}) is at least $CM^2qLn^{-1}$
%J_2 (\bgamma) \ge CM^2qLn^{-1}
uniformly on $\Gamma_M$ with probability tending to 1, where $C$
%Note that $C$ in (\ref{eqnh615})
does not depend on
$M$. Thus, we have %(\ref{eqnh611})-(\ref{eqnh615}) yield that
\[
\lim_{M\to\infty}\limsup_{n\to\infty} \RP\Bigl( \inf
_{\bgamma\in\Gamma_M}l_q( \bgamma)> l_q (
\bgamma_m ) \Bigr) =1. %
\]
It follows from the above result, the strict concavity of $l_q( \bgamma
)$, %(\ref{eqnh614})
and Lemma 6 %\ref{lemlem6}
of the supplementary material \cite{CHLP2014},
that there is a unique minimizer $\widetilde\bgamma$ of $l_q( \bgamma
)$ on $R^{(q+1)L}$ giving the desired convergence rate.
%Hence the proof of Theorem \ref{teothm3} is complete.

%%%%%%%%%%%%%%%%%%%%%%%%%%%%%%%%%%%%%%%%%%%%%%%%%% Theorem 4
%s6.2 #&#
\subsection{Proof of Theorem \texorpdfstring{\protect\ref{teothm4}}{3.2}}
Define $\overline\bgamma=(\overline\gamma_0^T,
\ldots, \overline\gamma_q^T)^T
\in R^{(q+1)L}$ by
%
%e6.4 #&#
\begin{equation}
\arg\min \bigl\| (\beta_0, \ldots, \beta_q)^T
- \bigl(\gamma_0^T\bB, \ldots, \gamma
_q^T\bB\bigr)^T \bigr\|_{L_2}^2.
\label{eqnh618}
\end{equation}
Then, $(\overline\gamma_0^T\bB, \ldots,
\overline\gamma_q^T\bB)^T \in\bG_0$ due to Assumption \ref{assS}(1), and
the minimum in (\ref{eqnh618}) is no larger than that of $\bgamma_m$
in Lemma 6 %\ref{lemlem6}
of the supplementary material \cite{CHLP2014}. Thus, Lemma 6 %\ref{lemlem6}
of the supplementary material \cite{CHLP2014} and (\ref{eqnh598}) together
imply that, with probability tending to 1,
$
|\overline\bgamma- \bgamma_m |^2 \le C \rho_{qn}^2 L
$
for some positive $C$. The desired result follows if we
show that
%
%e6.5 #&#
\begin{equation}
\lim_{M\to\infty}\limsup_{n\to\infty} \RP\Bigl( \inf
_{\bgamma\in\overline\Gamma_M} Q_q( \bgamma) > Q_q( \overline
\bgamma) \Bigr) = 1, \label{eqnh616}
\end{equation}
where
$
\overline\Gamma_M = \{ \bgamma\in R^{(q+1)L} | |\bgamma-
\overline\bgamma| = ML^{1/2}r_{qn} \}$.
Write
%
%e6.6 #&#
\begin{eqnarray}\label{eqnh617}
Q_q( \bgamma) - Q_q( \overline\bgamma)
&=& \bigl\{ l_q ( \bgamma) -l_q( \overline\bgamma)\bigr\}\nonumber
\\
&&{} + \Biggl[ \sum_{k=0}^q
\bigl\{ p_{\lambda_1}\bigl( \bigl|(g_k)_{c}\bigr| \bigr) +
p_{\lambda_2}\bigl( \bigl\|(g_k)_{f}\bigr\|_{L_2}
\bigr)\bigr\}%\nonumber
\\
&&\hspace*{17pt}{} - \sum_{k=0}^q
\bigl\{ p_{\lambda_1}\bigl( \bigl|(\overline g_k)_{c}\bigr|
\bigr) + p_{\lambda_2}\bigl( \bigl\|(\overline g_k)_{f}
\bigr\|_{L_2} \bigr)\bigr\} \Biggr],
\nonumber
\end{eqnarray}
where $g_k= \gamma_k^T \bB$ and
$\overline g_k= \overline\gamma_k^T \bB$, $k=0,1,\ldots,q$.
%As in the proof of Theorem \ref{teothm3},
We have
\begin{eqnarray}\label{eqnh619}
&& l_q( \bgamma) - l_q( \overline\bgamma) \nonumber
\\
&&\qquad  =  2(\bgamma- \overline\bgamma)^T \bigl\{ -\bigl\langle\bigl(
\bB^T \town^T\bigr)^T, \varepsilon\bigr
\rangle_n + \Sigma_n % \nip{(\bB^T \town^T)^T, (\bB^T \town^T)}
(\overline\bgamma-
\bgamma_m)\bigr\} %\nonumber\\
%& & \qquad
\\
&&\quad\qquad{} + (\bgamma- \overline
\bgamma)^T %\nip{(\bB^T \town^T)^T, (\bB^T \town^T)}
\Sigma_n (\bgamma-\overline\bgamma).
\end{eqnarray}
Lemmas 5 and 7 %\ref{lemlem5} and \ref{lemlem7}
of the supplementary material \cite{CHLP2014} imply that uniformly in
\mbox{$\bgamma\in\overline\Gamma_M$}, the first term in the right-hand
side of (\ref{eqnh619}) equals
$ML^{1/2}r_{qn}\*O_p((q/n)^{1/2}) + MLr_{qn}O_p(L^{-1})\rho_{n}= MO_p(r_{qn}^2)$.
% J_{3}(\bgamma) & = ML^{1/2}r_{qn}O_p((q/n)^{1/2})
% + MLr_{qn}O_p(L^{-1})\rho_{n}\label{eqnh621}%\\
%&
%= MO_p(r_{qn}^2).%\notag
By Lemma 5 %\ref{lemlem5}
of the supplementary material \cite{CHLP2014}, there is a positive constant $C$ such that the
second term in the right-hand side of (\ref{eqnh619})
is no less than $CM^2r_{qn}^2$,
%J_{4}(\bgamma) \ge CM^2r_{qn}^2
uniformly in $\bgamma\in\overline\Gamma_M$
and with probability tending to 1. Thus, %(\ref{eqnh621})-(
we have
%
%e6.7 #&#
\begin{equation}
\lim_{M\to\infty}\limsup_{n\to\infty} \RP\Bigl( \inf
_{\bgamma\in\overline\Gamma_M} %\{ J_{3}(\bgamma) + J_{4}(\bgamma) \}
\bigl\{l_q( \bgamma) -
l_q( \overline\bgamma)\bigr\} \ge CM^2r_{qn}^2/2
\Bigr) = 1. \label{eqnh625}
\end{equation}

Next, we consider the difference between the penalty terms.
Recall that $a_0$ appearing below comes from the SCAD
function in (\ref{eqnh409}). When $| \bgamma-
\overline\bgamma|= ML^{1/2}r_{pn}$ we have,
for $k=0,1,\ldots,q$ and sufficiently large $M$,
\begin{eqnarray*}
\bigl|\bigl(\gamma_k^T \bB\bigr)_{c}\bigr|,
%> a_0 \lambda_1 {\rm and }
\bigl|\bigl(\overline\bgamma_k^T \bB
\bigr)_{c}\bigr| &>& a_0 \lambda_1\quad\mbox{or}\quad\bigl|\bigl(\gamma_k^T \bB\bigr)_{c} \bigr| = o(
\lambda_1)\quad\mbox{and}\quad \bigl| \bigl(\overline\bgamma_k^T \bB\bigr)_{c}\bigr| =0,
\\
%
%and
%
\bigl\|\bigl(\gamma_k^T\bB\bigr)_{f}
\bigr\|_{L_2}, %> a_0 \lambda_2 {\rm and }
\bigl\|\bigl(\overline\gamma_k^T
\bB\bigr)_{f}\bigr\|_{L_2} &>& a_0
\lambda_2\quad\mbox{or}\quad \bigl\| \bigl(\gamma_k^T\bB
\bigr)_{f} \bigr\|_{L_2} = o(\lambda_2)\quad\mbox{and}
\\
&&\hspace*{52pt} \bigl\| \bigl(\overline\gamma_k^T \bB\bigr)_{f}
\bigr\|_{L_2} = 0.
\end{eqnarray*}
The above relations and the properties of the SCAD function
imply the second term of the right-hand side of (\ref{eqnh617}) is
nonnegative on $\overline\Gamma_M$.
Thus, (\ref{eqnh616}) follows from (\ref{eqnh619})
and (\ref{eqnh625}), and the proof of Theorem \ref{teothm4} is complete.

%%%%%%%%%%%%%%%%%%%%%%%%%%%%%%%%%%%%%%%%%%%%%%%%%% Theorem 5
%s6.3 #&#
\subsection{Proof of Theorem \texorpdfstring{\protect\ref{teothm5}}{3.3}}
We prove the sparsity property in a way similar to the former half of
the proof of Theorem~1 in \cite{WLH2008}.
%First we prove that any local minimizer of $Q_q(\bgamma)$ on
%$R^{(q+1)L}$, $\widehat\bgamma=(\widehat\gamma_0^T, \ldots, \widehat
%is the unique minimizer of $l_q(\bgamma)$ on $\overline\bG_0$, where $
%oracle space $\bG_0$.
%&(\widehat\gamma_0^T\bB, \ldots,
% \in\bG_0&\label{eqnh801}\\
%&\| (\widehat\gamma_0^T\bB, \ldots, \widehat\gamma_q^T\bB)^T
% - (\beta_0,\ldots, \beta_q)^T \|_{L_2} \le\eta_nr_{qn}&
%Next we derive the $L_2$ convergence rate of the oracle estimator.
%Finally we verify that any local minimizer satisfying
%(\ref{eqnh803}) has the property of (\ref{eqnh801}) with
%probability tending to 1.

First, let $\widehat\bgamma$ be a local minimizer of $Q_q(\bgamma)$
on $R^{(q+1)L}$ %, $ \widehat\bgamma$,
satisfying% (\ref{eqnh801}) and (\ref{eqnh803}).
%
%e6.8 #&#
%e6.9 #&#
\begin{eqnarray}
\bigl(\widehat\gamma_0^T\bB, \ldots, \widehat
\gamma_q^T\bB\bigr)^T &\in&\bG_0,
\label{eqnh801}
\\
\bigl\| \bigl(\widehat\gamma_0^T\bB, \ldots, \widehat
\gamma_q^T\bB\bigr)^T - (\beta_0,
\ldots, \beta_q)^T \bigr\|_{L_2} &\le&
\eta_nr_{qn}. \label{eqnh803}
\end{eqnarray}
Then consider $Q_q(\widehat\bgamma+\bdelta)$ for
$ \widehat\bgamma+\bdelta\in\overline\bG_0$, where $\overline\bG
_0$ is the subspace of $R^{(q+1)L}$ corresponding to the oracle space
$\bG_0$. When $| \bdelta|$
is small enough, we have the same value of the penalty term as that
for $ \widehat\bgamma$ due to the flatness of the SCAD function.
On the other hand, the local optimality of $\widehat\bgamma$ implies
that
$
Q_q(\widehat\bgamma+\bdelta) \ge Q_q(\widehat\bgamma)$.
Thus, there is a small neighborhood of $\widehat\bgamma$
in $ \overline\bG_0 $, $\Gamma_h$, such that
$
\inf_{\bgamma\in\Gamma_h} l_q( \bgamma) \ge
l_q (\widehat\bgamma)$.
This shows that $\widehat\bgamma$ is a local minimizer of
$l_q(\bgamma)$ on $\overline\bG_0$. Since $l_q(\bgamma)$
is strictly concave on $R^{(q+1)L}$ with probability tending
to 1, this $\widehat\bgamma$ must be the unique minimizer of
$l_q(\bgamma)$ on $\overline\bG_0$, denoted by $\widehat\bgamma
_0$. A similar argument can be found in \cite{FL2011}.

Next, we deal with the oracle estimator $\widehat\bgamma_0$. %, the
%unique minimizer of $l_q(\bgamma)$ on $\overline\bG_0$.
We neglect $x^{(s+1)}(t),
\ldots, x^{(q)}(t)$ and restrict $\overline\bG_0$
to $R^{(s+1)L}$. Besides, we define $\bgamma_m\in
R^{(s+1)L}$ and $\overline\bgamma\in\overline\bG_0
\subset R^{(s+1)L}$ similarly as in the proofs of Theorems \ref{teothm3}
and \ref{teothm4}. Then we have
\begin{eqnarray*}
&& l_s( \bgamma) - l_s( \overline\bgamma)
\\
&&\qquad  =  2(\bgamma- \overline\bgamma)^T \bigl\{ -\bigl\langle\bigl(
\bB^T \town^T\bigr)^T, \varepsilon\bigr
\rangle_n + \Sigma_n %\nip{(\bB^T \town^T)^T, (\bB^T \town^T)}
(\overline\bgamma-
\bgamma_m)\bigr\} %\\
%& & \qquad
+(\bgamma- \overline
\bgamma)^T %\nip{(\bB^T \town^T)^T, (\bB^T \town^T)}
\Sigma_n (\bgamma-\overline\bgamma),
\end{eqnarray*}
for $\bgamma\in R^{(s+1)L}$, where $ \town$ is defined with $x^{(s+1)}(t),
\ldots, x^{(q)}(t)$ removed.
Pro\-ceeding in the same way as in the proof of
Theorem \ref{teothm4}, we obtain
$
\lim_{M\to\infty}\limsup_{n\to\infty}
\RP( \inf_{\bgamma\in\widetilde\Gamma_M}
l_s(\bgamma) > l_s(\overline\bgamma) )$,
where $\widetilde\Gamma_M = \{
\bgamma\in\overline\bG_0 | |\bgamma-
\overline\bgamma| = ML^{1/2}r_{sn} \}$.
Thus, % the unique minimizer of {\color{red} $l_q(\bgamma)$ on $
$\widehat\bgamma_0$ satisfies
$| \widehat\bgamma_0 - \overline\bgamma|
= O_p(L^{1/2}r_{sn})$.

Finally,\vspace*{1pt} we consider a local minimizer $\widehat\bgamma$ satisfying
(\ref{eqnh803}) and prove that it also satisfies
(\ref{eqnh801}).
%Take a local minimizer $\widehat\bgamma$ satsifying $\| (\widehat
For this $\widehat\bgamma$, suppose that
$(\widehat\beta_{j})_c \ne0$ and $(\beta_{j})_c=0$
for some $j$, where $\widehat\beta_j = \widehat
\gamma_j^T \bB$.
Then we have that $0< |(\widehat\beta_{j})_c|=o (\lambda_1)$.
Define $\widehat\bbeta_t$ for $t \in[0,1/2]$ by
\[
\widehat\bbeta_t = \widehat\bbeta+ t (\widehat\bbeta_{-cj}
-\widehat\bbeta) = (1-t) \widehat\bbeta+ t \widehat\bbeta_{-cj},
\]
where $\widehat\bbeta= (\widehat\beta_0,
\ldots, \widehat\beta_q)^T$ and we define $ \widehat\bbeta_{-cj} $
by replacing $ \widehat\beta_{j} $ of $\widehat\bbeta$
with $(\widehat\beta_j)_f$.

Defining $\widehat\bgamma_t = (\widehat\gamma_{0t}^T,
\ldots, \widehat\gamma_{qt}^T )^T$ by
$
\widehat\bbeta_t = ( \widehat\gamma_{0t}^T\bB,
\ldots, \widehat\gamma_{qt}^T\bB)^T$,
we evaluate
\begin{eqnarray*}
Q_q (\widehat\bgamma_t ) - Q_q
(\widehat\bgamma) %
& = & \bigl\{ l_q(
\widehat\bgamma_t ) - l_q(\widehat\bgamma) \bigr\} +
\bigl\{ p_{\lambda_1}\bigl((1-t)\bigl| (\widehat\beta_{j})_c\bigr|
\bigr) - p_{\lambda_1}\bigl(\bigl| ( \widehat\beta_{j})_c
\bigr| \bigr)\bigr\}
\nonumber
\\
& = & J_5 + J_6.
\end{eqnarray*}
It is easy to see that for some $\overline t \in[0,t]$,
$
J_6 = -t |(\widehat\beta_{j})_c| p_{\lambda_1}'(
(1-\overline t)|(\widehat\beta_{j})_c | )$.
%We deal with $J_5$.
In addition, we can represent $J_5$ as
\[
J_5 = - 2(\widehat\bgamma_t - \widehat
\bgamma)^T \Biggl\langle\bigl(\bB^T \town^T
\bigr)^T, y- \widehat\gamma_0^T\bB- \sum
_{k=1}^q\widehat\gamma_k^T
\mathbf{W}_{k} \Biggr\rangle_n +(\widehat
\bgamma_t - \widehat\bgamma)^T \Sigma_n (
\widehat\bgamma_t -\widehat\bgamma). %
\]
Lemmas 5 and 7\vspace*{1pt} %\ref{lemlem5} and \ref{lemlem7}
of the supplementary material \cite{CHLP2014} imply that the two terms in~$J_5$ can be
expressed\vspace*{2pt} as $ -2(\widehat\bgamma_t - \widehat\bgamma)^T
\langle(\bB^T \town^T)^T, (\bB^T \town^T) \rangle_n
(\bgamma_m - \widehat\bgamma) - 2(\widehat\bgamma_t - \widehat
\bgamma)^T
\langle(\bB^T \town^T)^T, \varepsilon\rangle_n$ and $t|(\widehat
\beta_{j})_c | O_p (\eta_n r_{qn})$.
%J_{51}& = -2(\widehat\bgamma_t - \widehat\bgamma)^T
%(\bgamma_m - \widehat\bgamma)\\
%& \qquad- 2(\widehat\bgamma_t - \widehat\bgamma)^T
%& =
%t|( \widehat\beta_{j})_c |O_p (r_{qn}), \\
%J_{52}& = t^2 |(\widehat\beta_{j})_c|^2 O_p(1) =
%t|(\widehat\beta_{j})_c | O_p (\eta_n r_{qn}).
Hence we get
$
J_5 = t| (\widehat\beta_{j})_c | O_p\* (\eta_n r_{qn})$.
From the above results, %(\ref{eqnh631})-(\ref{eqnh635}),
property of the SCAD function and Assumption~\ref{assS}(2),
\[
Q_q ( \widehat\bgamma_t ) - Q_q ( \widehat
\bgamma) = t\bigl| (\widehat\beta_{j})_c \bigr| \bigl\{
O_p(\eta_n r_{qn}) - p_{\lambda_1}'
\bigl( (1 - \overline t )\bigl|(\widehat\beta_{j})_c\bigr| \bigr)
\bigr\} < 0
\]
uniformly in $t\in(0,1/2)$ with probability tending to
1, and the probability does not depend on the specific value of $j$.
This contradicts with the local optimality of~$\widehat\bgamma$,
and thus $(\widehat\beta_{j})_c$ must be equal to $0$ if
$(\beta_{j})_c=0$. We can treat the other cases in the same way.
Hence, (\ref{eqnh801}) is established for the local minimizer
$\widehat\bgamma$, and the proof is complete.
\end{appendix}

\begin{supplement}%[id=suppA]\label{suppA}
\sname{Some technical material}
\stitle{Supplement to ``Nonparametric independence screening and
structure identification for ultra-high dimensional longitudinal data''}
\slink[doi]{10.1214/14-AOS1236SUPP} %[doi,text={...}] - jei reikia
%suskaldyti doi
\sdatatype{.pdf}
\sfilename{aos1236\_supp.pdf}
\sdescription{Some lemmas, and proofs of Theorems \ref{teothm1}--\ref
{teothm2} and Remark \ref{remarkrmk1}.}
\end{supplement}


%suskaldyti doi
% imsref loaded by linak, 2014-06-04 15:46:26
%

\printaddresses
\end{document}